\documentclass[twocolumn,aps,prd,nofootinbib]{revtex4-1}
\usepackage{graphicx}
\usepackage{hyperref}
\usepackage{xcolor}

\graphicspath{{./fig/}{./png/}}
\usepackage{amsmath}
\newcommand{\EQ}{\begin{equation}}
\newcommand{\EN}{\end{equation}}
\newcommand{\EQA}{\begin{eqnarray}}
\newcommand{\ENA}{\end{eqnarray}}

\newcommand{\Eq}[1]{Eq.~(\ref{#1})}

\newcommand{\Sec}[1]{Section~\ref{#1}}

\newcommand{\Fig}[1]{Fig.~\ref{#1}}

\newcommand{\Figp}[2]{Fig.~\ref{#1}({#2})}
\newcommand{\Figsp}[3]{Figs.~\ref{#1}({#2}) and ({#3})}
\newcommand{\Figs}[2]{Figs.~\ref{#1} and \ref{#2}}

\newcommand{\Tab}[1]{Table~\ref{#1}}

\newcommand{\App}[1]{Appendix~\ref{#1}}

\newcommand{\bra}[1]{\langle #1\rangle}

{}
{}

\DeclareMathAlphabet\mathbfcal{OMS}{cmsy}{b}{n}

\newcommand{\kk}{\mathbf{k}}

\newcommand{\qq}{\mathbf{q}}

\newcommand{\BB}{\mathbf{B}}

\newcommand{\JJ}{\mathbf{J}}

\newcommand{\AAA}{\mathbf{A}}

\newcommand{\uu}{\mathbf{u}}

\newcommand{\nab}{{\mathbf{\nabla}}}

\newcommand{\SSSS}{\mbox{\boldmath ${\sf S}$} {}}

\newcommand{\hh}{\mbox{\boldmath ${\sf h}$} {}}
\newcommand{\TT}{\mbox{\boldmath ${\sf T}$} {}}

\newcommand{\TTT}{{\sf T}}
\newcommand{\tildeh}{\tilde{h}}
\newcommand{\tildeT}{\tilde{T}}

\newcommand{\DD}{{\rm D} {}}

\newcommand{\dd}{{\rm d} {}}

\def\la{\mathrel{\mathchoice {\vcenter{\offinterlineskip\halign{\hfil
$\displaystyle##$\hfil\cr<\cr\sim\cr}}}
{\vcenter{\offinterlineskip\halign{\hfil$\textstyle##$\hfil\cr<\cr\sim\cr}}}
{\vcenter{\offinterlineskip\halign{\hfil$\scriptstyle##$\hfil\cr<\cr\sim\cr}}}
{\vcenter{\offinterlineskip\halign{\hfil$\scriptscriptstyle##$\hfil\cr<\cr\sim\cr}}}}}

\def\H{\mbox{\rm H}}

\def\Sp{\mbox{\rm Sp}}

\def\EEM{{\cal E}_{\rm M}}
\def\EEEM{{\cal E}_{\rm EM}}

\def\EEGW{{\cal E}_{\rm GW}}
\def\OmGW{{\Omega}_{\rm GW}}

\def\EM{E_{\rm M}}

\def\te{t_{\rm e}}

\def\kH{k_{\rm H}}

\def\kp{k_{\rm p}}

\def\EM{E_{\rm M}}

\def\half{{\textstyle{1\over2}}}

\def\onethird{{\textstyle{1\over3}}}

\def\power{p}

\def\runh{\text{HEL}}
\def\runnh{\text{NHEL}}
\def\apjl{Astrophys. J. Lett.}
\def\jcap{J. Cosmol. Astropart. Phys.}
\def\mnras{Mon. Not. Roy. Astron. Soc.}
\begin{document}


\title{Low frequency tail of gravitational wave spectra from hydromagnetic turbulence}
\author{Ramkishor Sharma$^{1,2}$}
\email{ramkishor.sharma@su.se}
\author{Axel Brandenburg$^{1,2,3,4}$} 
\email{brandenb@nordita.org}
\affiliation{$^{1}$Nordita, KTH Royal Institute of Technology and Stockholm University,
Hannes Alfv\'ens v\"ag 12, 10691 Stockholm, Sweden}
\affiliation{$^{2}$The Oskar Klein Centre, Department of Astronomy,
Stockholm University, 10691 Stockholm, Sweden}
\affiliation{$^{3}$McWilliams Center for Cosmology \& Department of Physics,
Carnegie Mellon University, Pittsburgh, PA 15213, USA}
\affiliation{$^{4}$School of Natural Sciences and Medicine, Ilia State University,
3-5 Cholokashvili Avenue, 0194 Tbilisi, Georgia}

\begin{abstract}
Hydrodynamic and magnetohydrodynamic (MHD) turbulence in the early
Universe can drive gravitational waves (GWs) and imprint their spectrum
onto that of GWs, which might still be observable today.
We study the production of the GW background from freely decaying MHD
turbulence from helical and nonhelical initial magnetic fields.
To understand the produced GW spectra, we develop a simple model on
the basis of the evolution of the magnetic stress tensor.
We find that the GW spectra obtained in this model reproduce those
obtained in numerical simulations if we consider the detailed time evolution of
the low frequency tail of the stress spectrum from numerical simulations.
We also show that the shapes of the produced GW frequency spectra are different
for helical and nonhelical cases for the same initial magnetic energy spectra.
Such differences can help distinguish helical and nonhelical
initial magnetic fields from a polarized background of GWs --
especially when the expected circular polarization cannot
be detected directly.

\end{abstract}

\maketitle
\section{Introduction}

Magnetohydrodynamic (MHD) turbulence in the early Universe can be a powerful source
of gravitational waves (GWs) that could be observable as a stochastic
background today \citep{1994PhRvD..49.2837K,PhysRevLett.85.2044,2002PhRvD..66b4030K,Dolgov+02,2018CQGra..35p3001C}.
The frequency spectrum of these waves is related to the spectrum of the
underlying turbulence.
Such turbulence could be induced during the various epochs\footnote{The electroweak
and QCD epochs are accompanied by crossovers in the standard model
\cite{PhysRevLett.77.2887,2006Natur.443..675A}. However, many extensions of the
standard model can lead to a first order phase transition.} in the early
Universe \citep{PhysRevD.30.272, PhysRevD.30.272,10.1093/mnras/218.4.629,
Mazumdar_2019,10.21468/SciPostPhysLectNotes.24} or the possible presence
of primordial magnetic fields \citep{1988PhRvD..37.2743T, tanmay1991,
1992ApJ...391L...1R,durrer2013, subramanian2016, kandu2019, tanmay2021,
Bran+He+Shar21}.
These GWs produced by turbulence at an epoch of the electroweak phase
transition lie in the sensitivity range of the proposed
Laser Interferometer Space Antenna and pulsar timing arrays for
the turbulence induced around an epoch of the quantum chromodynamics
(QCD) phase transition.
Recently, various pulsar timing arrays \citep{NANOGrav2020,
Goncharov_2021, Chen2021, Antoniadis2022} have reported evidence for
the presence of a common spectrum process across analyzed pulsars in
the search of the presence of an isotropic stochastic GW background.
This evidence has been used to constrain the strength and
correlation length of magnetic fields generated at the QCD
epoch \citep{2021PhRvD.103L1302N, Sharma21, RoperPol+22}.
However, the presence of a quadrupolar spatial correlation
\citep{Hellings&Downs}, a characteristics of a GW background,
is yet to be claimed.

Numerical simulations have confirmed that there is indeed a direct
connection between the slopes of the turbulence and GW spectra
\citep{RoperPol+20}, except that at low frequencies, below the peak of
the spectrum, the GW spectrum was found to be shallower in the simulations
than what was previously expected from analytical calculations.
We call this part the low frequency tail of the GW spectrum.
However, there is the worry that this shallow tail could be caused by
unknown numerical artifacts such as the finite size of the computational
domain and the way the turbulence is initiated in the simulations.

To understand the origin of the low frequency tail,
the authors of Ref.~\cite{RoperPol+22} have recently compared numerical
MHD simulations with an analytic model, where the stress is assumed constant
for a certain interval of time. Their model predicts a flat spectrum whose extent depends on the duration
over which the stress is held constant.
In this way, it was possible to determine an effective duration for a
given numerical simulation.
This duration was found to be different for different simulations.
Their model is therefore descriptive rather than predictive.
Furthermore, in the numerical solutions the stress was
not actually constant for any duration of time.

In another recent approach, the authors of Ref.~\cite{Auclair:2022jod} have focused on the
importance of unequal time correlation functions of the Fourier components
of the velocity field for purely hydrodynamic turbulence.
While the authors acknowledge the potential importance of the initial
growth phase of the turbulence, they also have no inverse
cascade in their simulations.
This is different from MHD turbulence, which can display inverse cascading
even in the absence of net magnetic helicity \cite{2015PhRvL.114g5001B,2014ApJ...794L..26Z}.
This will be crucial to the approach discussed in the present paper.

To address the problem of a limited computational domain, it is important to use large enough computational domains so that its minimum
wave number is as small as possible.
In this paper, we discuss two MHD simulations where the wave number
corresponding to the peak of the GW spectrum and the wave numbers below
that corresponding to the horizon size at the initial time are well
resolved.
Since the stress appears explicitly in the linearized GW
equation, we also analyze for these simulations the evolution of the stress spectrum along with the magnetic and GW spectra.
Such a detailed comparison between the simulated stress and
resulting GWs is an important new aspect of the present work.
Second, we develop a simple model, motivated by the stress evolution seen
in the present simulations, to explain the GW spectrum obtained.
In this model, our main focus is to understand the nature of the GW
spectrum below the wave number corresponding to the peak of the spectrum.
Our simulations are similar to those of Ref.~\cite{RoperPol+22},
but our interpretation and corresponding modeling of the stress is not.
There is no time interval during which the stress is constant.
Our results are therefore not characterized by the duration
of such a time interval.
In addition, we determine and analyze spectral differences between runs with
and without magnetic helicity, which were not noticed previously.
We also emphasize that the Hubble horizon wave number poses an ultimate
cutoff for the flat spectrum toward low wave numbers.

This paper is organized as follows.
In \Sec{TheModel}, we discuss the evolution of the magnetic field, stress,
and GW spectrum in our new runs.
In this section, we also discuss how the stress spectrum evolves when
inverse transfer and inverse cascade of the turbulence correspond to
the evolution for nonhelical and helical magnetic fields in the early
Universe.
In \Sec{simplemodel}, we discuss the model to explain the low frequency
tail of the GW spectrum.
Further, in \Sec{comparison}, we compare the GW spectrum obtained from
our numerical simulations and our model.
We conclude in \Sec{conclusion}.

\section{Nonhelical and helical cascades}
\label{TheModel}

Various phenomena such as primordial magnetic fields and phase
transitions can lead to the generation of turbulence in the early
Universe.
The stress associated with magnetic fields and turbulence
lead to the production of GWs.
This has been studied in the literature both analytically \citep{Dolgov+02,kosowsky2002,Gogo+07,tina2008,Chiara2009,sigl2018,Sharma+20}
and numerically \citep{RoperPol+20,RoperPol+21,Kahniashvili+21,Jani2021,RoperPol+22,Auclair:2022jod}.
In the present paper, we perform new simulations of decaying
MHD turbulence, where we resolve the scales which are smaller than
the Hubble horizon size at the initial time.
Before explaining the simulations in detail, let us begin by summarizing
the basic equations.

\subsection{GWs from MHD turbulence}
\label{GWfromCME}

We follow here the formalism of Ref.~\cite{RoperPol+20b, RoperPol+20}, where
conformal time is normalized to unity at the initial time.
One could associate this with the electroweak phase transition,
for example.
The velocity $\uu$ is normalized to the speed of light.
The magnetic field $\BB=\nab\times\AAA$ is written in terms of the
magnetic vector potential $\AAA$, and the current density is written
as $\JJ=\nab\times\BB$.
Following Ref.~\cite{BEO96}, the energy density $\rho$ includes the restmass
density, so its evolution equation obeys a continuity equation that also
includes magnetic energy terms.
As in \cite{RoperPol+20b}, $\rho$ is normalized to the critical energy
density for a flat Universe.
We solve for the Fourier transformed plus and cross polarizations
of the gravitational strain, $\tilde{h}_+$ and $\tilde{h}_\times$,
which are driven by the corresponding projections of the stress,
which, in turn, is composed of kinetic and magnetic contributions,
\EQ
\TTT_{ij} =\frac{4}{3}\gamma_{\rm Lor}^2\rho u_i u_j-B_i B_j+...,
\EN
where $\gamma_{\rm Lor}=(1-\uu^2)^{-1/2}$ is the Lorentz
factor, and the ellipsis denotes terms proportional to $\delta_{ij}$,
which do not contribute to the projected source $\tilde{T}_{+/\times}$.

Assuming the Universe to be conformally flat, its expansion can be
scaled out by working with conformal time $t$ and comoving variables
\citep{BEO96}.
We use the fact that in the radiation-dominated era, the scale factor
grows linearly with conformal time.
The only explicit occurrence of conformal time is then in the GW
equation, where a $6/t$ factor occurs in the source term \citep{RoperPol+20b}.
The full set of equations is therefore
\begin{eqnarray}
&&\frac{\partial\BB}{\partial t}=
\nab\times(\uu\times\BB-\eta\nab\times\BB),\label{dAdt}\\
&&{\DD\uu\over\DD t}=
{1\over\rho}\nab\cdot\left(2\rho\nu\SSSS\right)-{1\over4}\nab\ln\rho
+{\uu\over3}\left(\nab\cdot\uu+\uu\cdot\nab\ln\rho\right)
\nonumber \\
&&\qquad\quad-{\uu\over\rho}
\left[\uu\cdot(\JJ\times\BB)+\eta \JJ^2\right]
+{3\over4\rho}\JJ\times\BB,
\label{dudt} \\
&&{\partial\ln\rho\over\partial t}
=-\frac{4}{3}\left(\nab\cdot\uu+\uu\cdot\nab\ln\rho\right)
+{1\over\rho}\left[\uu\cdot(\JJ\times\BB)+\eta \JJ^2\right]\!,
\nonumber \\
&&\frac{\partial^2}{\partial t^2} \tilde{h}_{+/\times} (\kk, t) 
+k^2\tilde{h}_{+/\times} (\kk, t) = {6\over t} 
\tilde{T}_{+/\times}(\kk,t),
\label{GW4}
\end{eqnarray}
where $\DD/\DD t\equiv\partial/\partial t+\uu\cdot\nab$ is the advective derivative,
$\eta$ is the magnetic diffusivity, $\nu$ is the kinematic viscosity,
${\sf S}_{ij}=\half(u_{i,j}+u_{j,i})-\onethird\delta_{ij}\nab\cdot\uu$ are
the components of the rate-of-strain tensor $\SSSS$ with commas denoting
partial derivatives.
Fourier transformation in space is denoted by a tilde.
In all cases studied in this paper, the initial conditions are such
that $\BB$ consists of a weak Gaussian-distributed seed magnetic field,
$\uu=0$, $\rho=1$.

We work with spectra that are defined as integrals over concentric shells
in wave number space $\kk$ with $k=|\kk|$.
They are normalized such that their integrals over $k$ give the mean square
of the corresponding quantity, i.e., $\int\Sp(\BB)\,\dd k=\bra{\BB^2}$,
where $\Sp(\BB)=\Sp(B_x)+\Sp(B_y)+\Sp(B_z)$.
Similarly, $\Sp(\hh)=\Sp(h_+)+\Sp(h_\times)$ is defined as the sum over
the two polarization modes.
Of particular interest will also be the stress spectrum $\Sp(\TT)$,
which is defined analogously through $\Sp(\TT)=\Sp(T_+)+\Sp(T_\times)$.
To study the evolution of the stress at selected Fourier modes, we compute
$|\tildeT(k,t)|\equiv\sqrt{\Sp(\TT)/4\pi k^2}$, which scales the same
way as $|\tilde{T}_+(k,t)|$ and $|\tilde{T}_\times(k,t)|$.

\subsection{Evolution of the stress and strain spectra}
\label{EvolutionOfStress}

To put our results into perspective and compare with earlier work,
we study cases of suddenly initiated turbulence.
We perform simulations similar to those of 
Ref.~\cite{RoperPol+20}
by using as
initial condition for the magnetic field a random Gaussian-distributed
magnetic field with a $k^4$ spectrum for $k<k_{\rm p}$ and a $k^{-5/3}$
spectrum for $k>k_{\rm p}$.
For details of such a magnetic field, see Ref.~\cite{Bran+17}.
As initial condition for the GW field, we assume that $h$ and $\dot{h}$
vanish.
The strength of the GW field is then strongly determined by the
sudden initialization of a fully developed turbulence spectrum.
The details of the simulations are given in \Tab{table1}.
In this table, the first column represents the name of the runs (HEL and NHEL),
$\EEEM^{i}$ is the initial value of the magnetic energy density compared
to the background energy density, $\kp$ is the wave number at which the
magnetic energy spectrum peaks and it is normalized by the wave number
corresponding to the Hubble horizon size at the initial time,
$\EEGW^{\rm sat}$ is the value of the GW energy density after
saturation compared to the background energy density, and
$\OmGW^{\rm sat}$ is the density parameter of GWs, representing the
ratio of the GW energy density compared to the critical energy density
at present.
$\OmGW^{\rm sat}$ has been calculated considering the production of GWs
around the electroweak phase transition.

\begin{figure*}\begin{center}
\includegraphics[width=\textwidth]{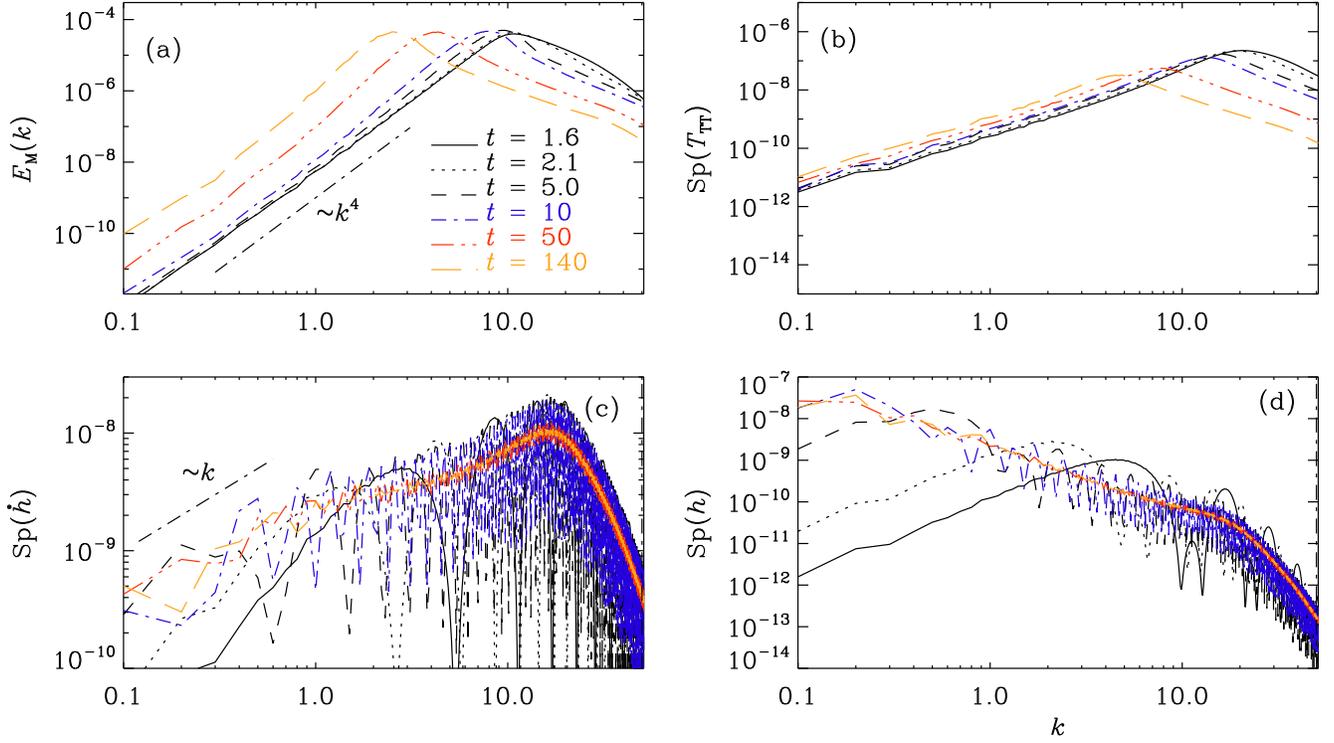}
\end{center}\caption{
Spectra of the magnetic field, the TT-projected stress, the strain
derivative, and the strain for suddenly initiated turbulence with
magnetic helicity.
}\label{pstress_etc_LowFreq_sig1}\end{figure*}

\begin{figure*}\begin{center}
\includegraphics[width=\textwidth]{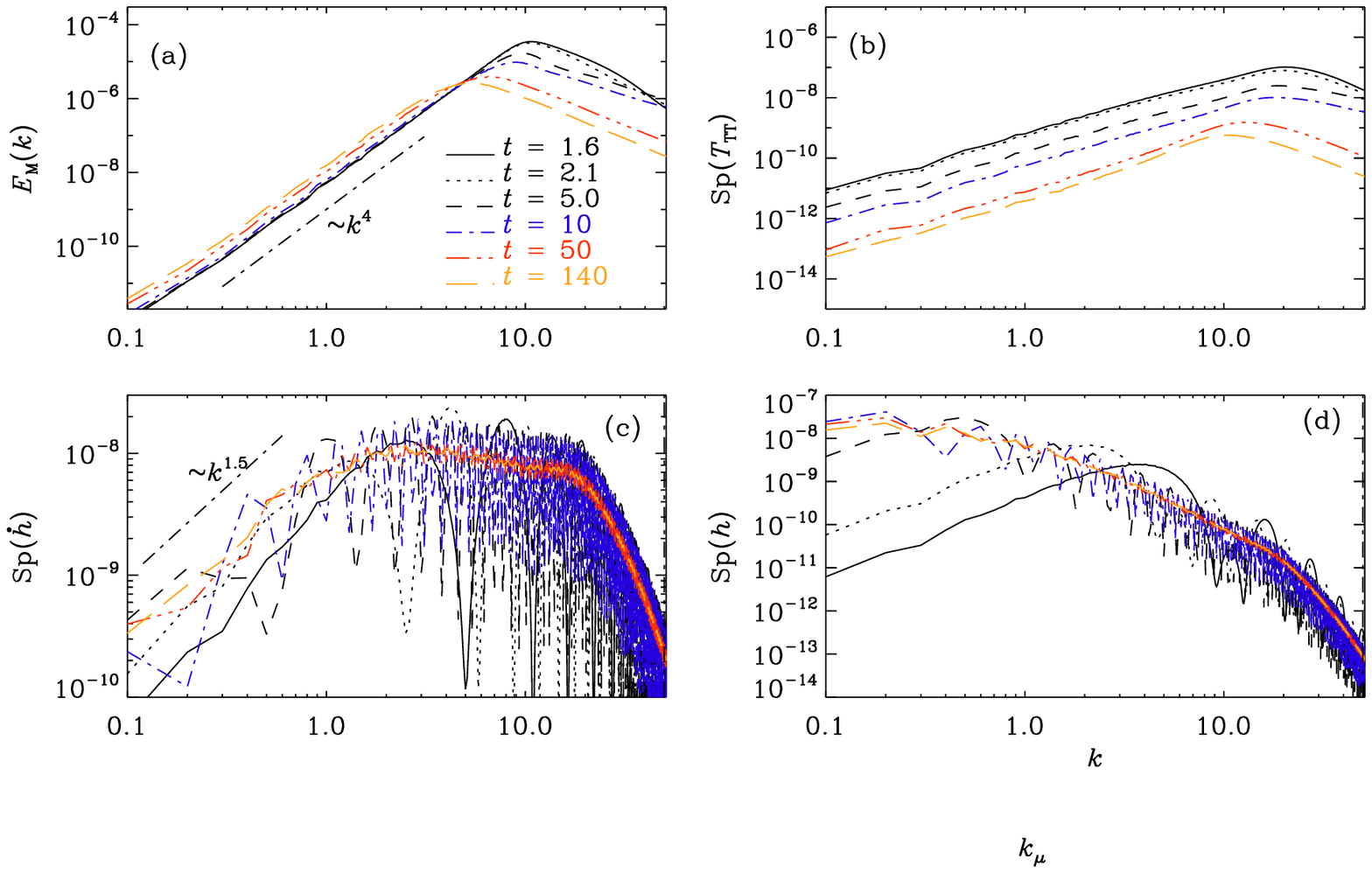}
\end{center}\caption{
Same as \Fig{pstress_etc_LowFreq_sig1}, but for the nonhelical case.
}\label{pstress_etc_LowFreq_sig0}\end{figure*}

We consider two runs where the initial magnetic field is either
fully helical ($\runh$) or nonhelical ($\runnh$).
In \Figs{pstress_etc_LowFreq_sig1}{pstress_etc_LowFreq_sig0}, we
show for HEL and NHEL, respectively, the time evolution of
the spectra of the magnetic field, the TT-projected stress, the strain
derivative, and the strain.
Inverse cascading is seen in the magnetic energy spectra, which leads
to the expected increase of the spectral stress at small $k$; see
\Figsp{pstress_etc_LowFreq_sig1}{a}{b} for the 
$\runh$ run.
By contrast, in $\runnh$, the stress spectrum always decreases at small $k$.
We also see in \Figp{pstress_etc_LowFreq_sig1}{c} that the GW energy
spectrum has a maximum at $k\approx20$, which is not present in $\runnh$;
cf.\ \Figp{pstress_etc_LowFreq_sig0}{c}.
Their spectra fall off toward smaller $k$ proportional to $k$ and
$k^{1.5}$ for $\runh$ and $\runnh$, respectively. 

\begin{table}\caption{
Summary of simulation parameters
}\begin{center}
\begin{tabular}{ccccccc}
\hline
Run & $\EEM^i$ &$\kp$ & $\EEGW^{\rm sat}$ & $\OmGW^{\rm sat}$\\
\hline
\runh & $5.4\times10^{-3}$ &$10$& $3.7\times10^{-7}$ &  $5.9\times10^{-12}$\\ 
\hline
\runnh & $5.5 \times 10^{-3}$ &$10$& $3.5\times10^{-7}$ &  $5.6\times10^{-12}$\\
\hline
\label{table1}\end{tabular}
\end{center}
\end{table}

\begin{figure*}\begin{center}
\includegraphics[width=\textwidth]{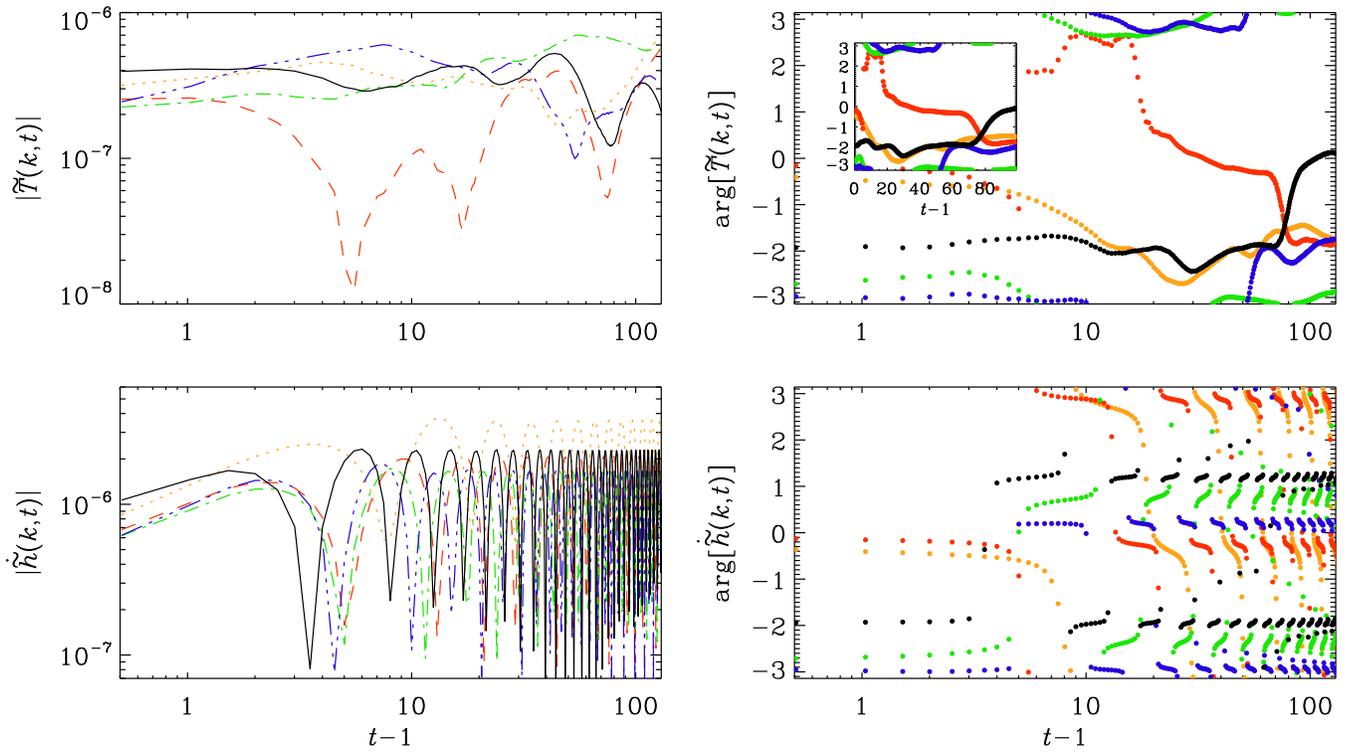}
\end{center}\caption{
Modulus and phase of $\tilde{T}(k,t)$ and $\dot{\tilde{h}}(k,t)$
for the helical case for $\kk=(k,0,0)$ with $k=0.3$ (orange), 0.4 (red),
0.5 (green), 0.6 (blue), and 0.7 (black).
The inset shows the phase with a linear abscissa.
}\label{rslice_stress_plot_LowFreq}\end{figure*}

\begin{figure*}\begin{center}
\includegraphics[width=\textwidth]{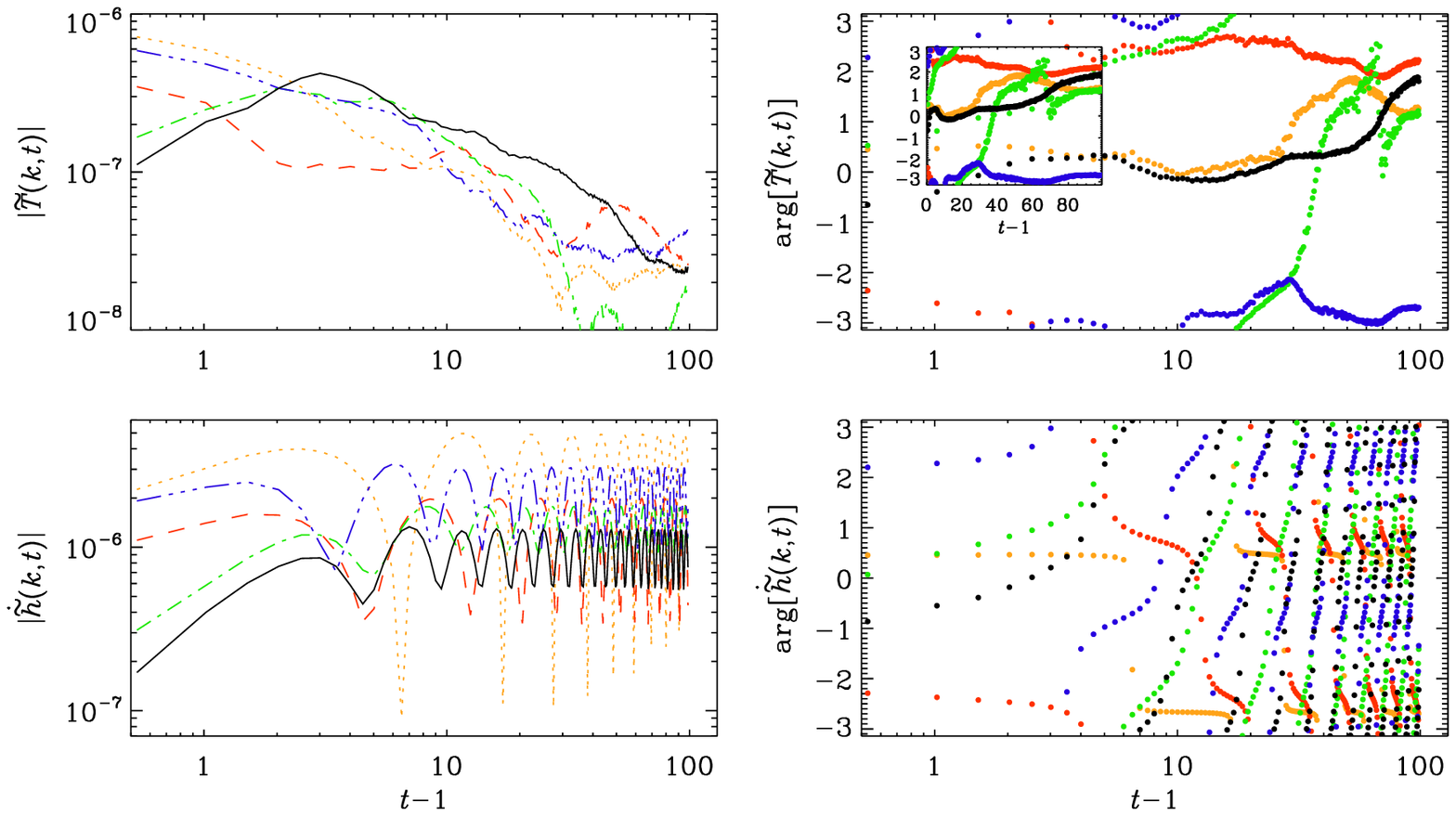}
\end{center}\caption{
Same as \Fig{rslice_stress_plot_LowFreq}, but for the nonhelical case.
}\label{rslice_stress_plot_LowFreq_nohel}\end{figure*}

\begin{figure*}\begin{center}
\includegraphics[scale=0.7]{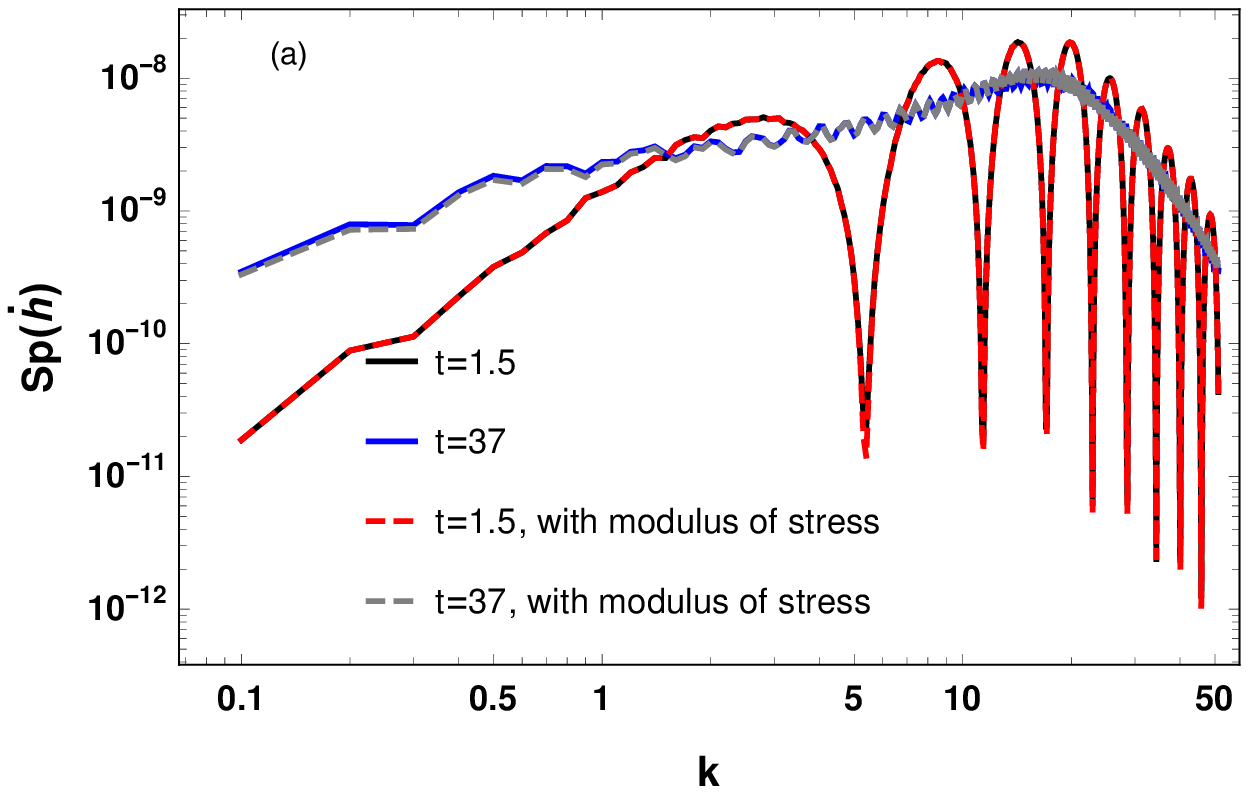}
\includegraphics[scale=0.7]{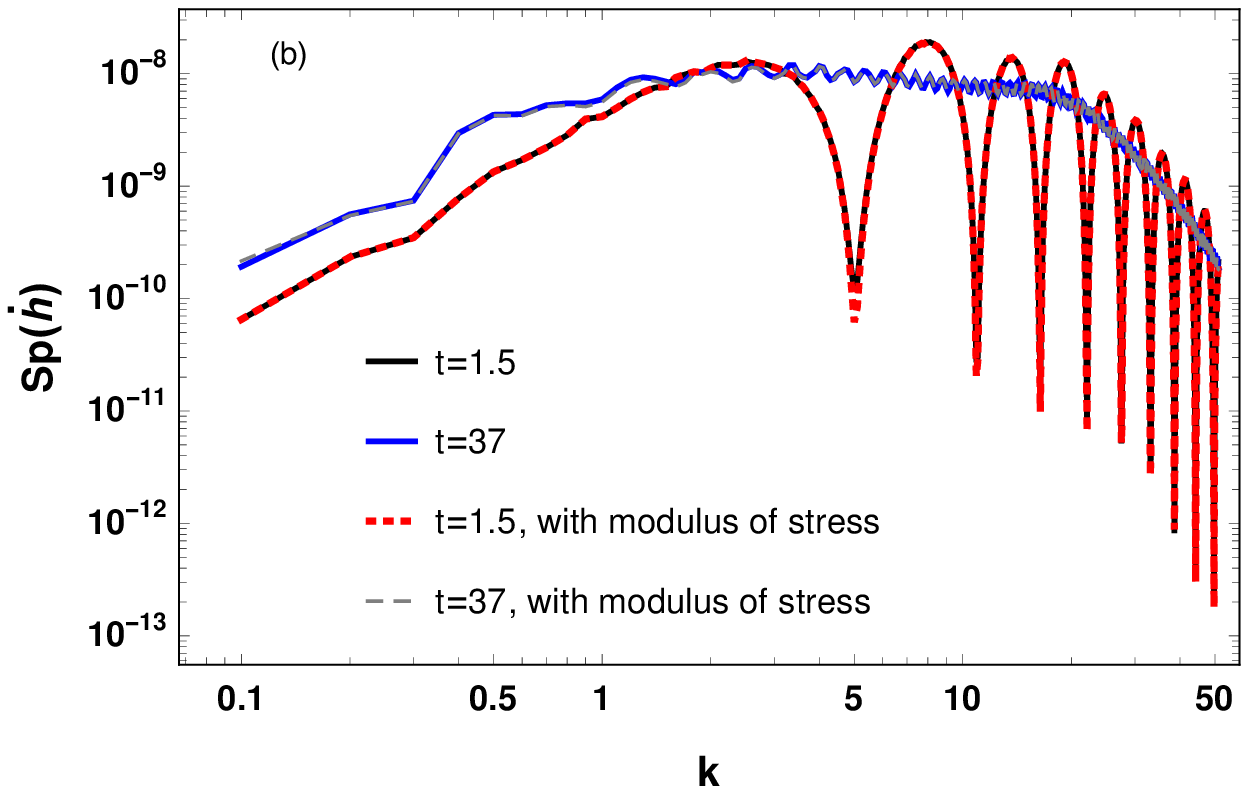}
\end{center}\caption{$\Sp(\dot{\hh})(k,t)$ vs $k$:
(a) The solid black and blue curves represent $\Sp(\dot{\hh})(k,t)$
at times $t=1.5$ and $t=37$ for run hel.
The dashed red and gray
curves show $\Sp(\dot{\hh})(k,t)$ for
the case when the stress spectrum has been replaced by its modulus
in the GW evolution equation. The black curve coincides with the dashed red curve and solid blue curve
coincides with dashed gray curve.
(b) Same as (a), but for run nonhel.}
\label{with_and_wo_phase}
\end{figure*}

In \Figs{rslice_stress_plot_LowFreq}{rslice_stress_plot_LowFreq_nohel},
we show the time evolution of the modulus and phase of stress and
strain derivative for $\runh$ and $\runnh$ for wave vectors $\kk=(k,0,0)$
and five different values of $k$, which are all below $k_{\rm p}$.
The purpose of this is to see how representative these individual
wave vectors are compared to the collective effect of all others
of similar length, and whether there is any important effect
resulting from the phases of the stress.

Broadly speaking, the time evolution of the modulus of the stress $|\tildeT|$ for
any of the five wave vectors does not seems to reflect the expectation
from the evolution of the shell-integrated stress spectrum, which is increasing
for $\runh$ and decreasing for $\runnh$, as was see in
\Figs{pstress_etc_LowFreq_sig1}{pstress_etc_LowFreq_sig0}.
This is an important observation and may be due to the fact that in
\Figs{rslice_stress_plot_LowFreq}{rslice_stress_plot_LowFreq_nohel},
we have shown stress and strain derivatives only for
particular values of $k$.

From \Figs{rslice_stress_plot_LowFreq}{rslice_stress_plot_LowFreq_nohel},
it is evident that $\arg(\tildeT)$ remains constant
for some time and starts evolving more rapidly after that.
It is also interesting to note that the amplitude of $|\dot{\tildeh}|$
increases up to the time until which $\arg(\dot{\tildeh})$
is roughly constant.
After this time, $|\dot{\tildeh}|$ enters an oscillatory regime
and its amplitude does not change much.
Other wave vectors of the same length show a similar behavior of
$\arg(\tildeT)$ and $\arg(\dot{\tildeh})$, as is shown in the figures of
the Supplemental Material provided along with the data in Ref.~\cite{DATA}.
This conclusion applies for both runs shown in
\Figs{rslice_stress_plot_LowFreq}{rslice_stress_plot_LowFreq_nohel} and
it leads us to develop a simple model to understand the GW spectrum in
these cases.
In this model we replace $\tildeT$ by its wave vector-averaged
magnitude, $|\tildeT|$, as discussed in \Sec{simplemodel}.

Further, to understand the role of the phases of the stress tensor in
the production of GWs, we run two new simulations analogous to 
runs $\runh$ and $\runnh$, where we replace $\tildeT(k,t)$ by
its modulus at each time step.
The final GW spectrum in these modified runs turns out to be
virtually the same as in the original $\runh$ and $\runnh$ runs;
see \Fig{with_and_wo_phase}.
The comparisons of HEL and NHEL are shown in
panels (a) and (b) of this figure, respectively.
Dashed red and gray curves at times $t=1.5$ and $t=37$, respectively,
are for the cases when $\tildeT(\kk,t)$ has been replaced by its modulus.
It is evident from the figure that there is hardly any difference in the
actual $\Sp(\dot{\hh})$ after replacing
the stress with its modulus.
On the basis of this observation, we develop a model to obtain the GW
spectrum from the time evolution of the spectrum of the stress tensor.

\begin{figure}
\begin{center}
\includegraphics[scale=0.6]{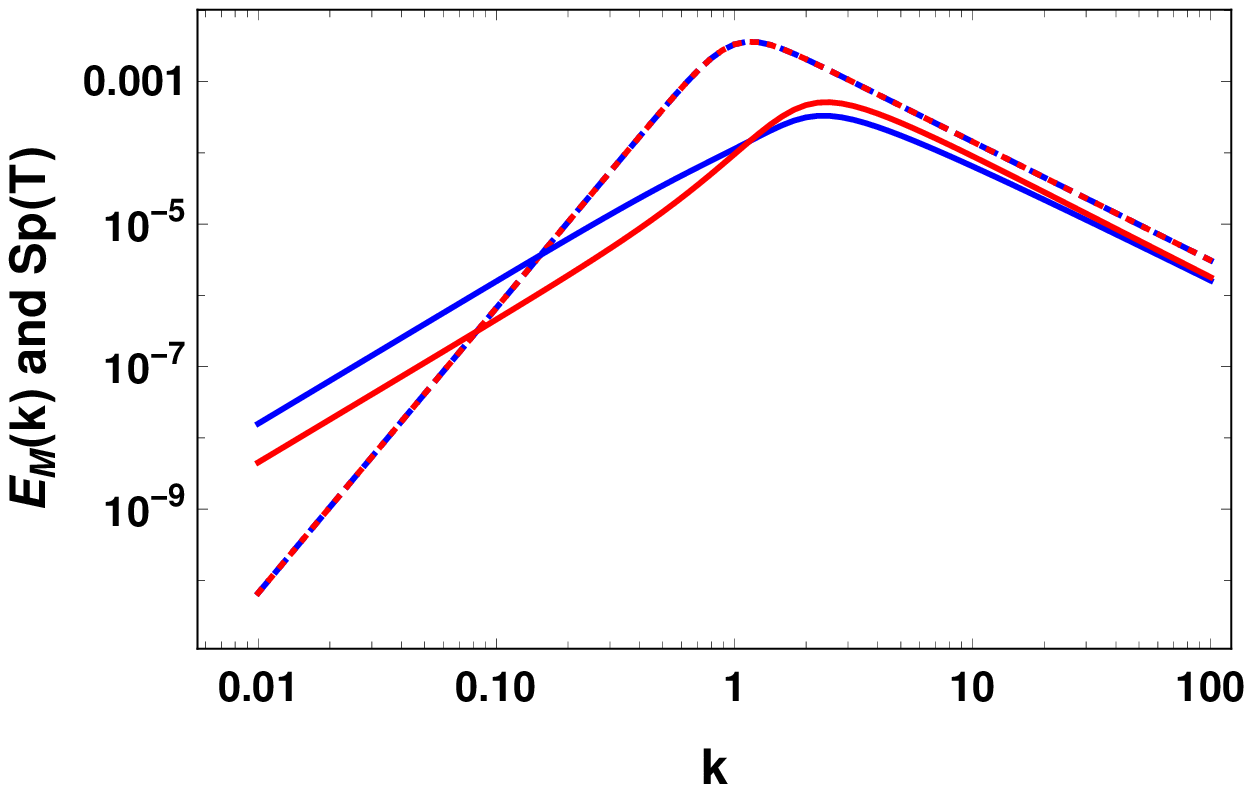}
\end{center}\caption{
In this figure, magnetic field energy spectrum, $\EM(k)$ (Dashed curves)
and $\Sp(\TT)$ (Solid curves) for the helical and non helical case.
The blue and red curves are for nonhelical and helical case respectively.
}\label{stress_spectrum_helvsnonhel}
\end{figure}

A striking difference between $\runh$ and $\runnh$ is the more
pronounced peak in the spectral GW energy in the former.
As we show in \App{appendixa}, this is due to the fact that the
stress spectrum for NHEL is different from that of HEL due to
the presence of additional helical contributions to the two-point correlation
of the magnetic field vectors.
This difference is shown in \Fig{stress_spectrum_helvsnonhel} and the
details are explained next.

\begin{figure*}\begin{center}
\includegraphics[scale=0.7]{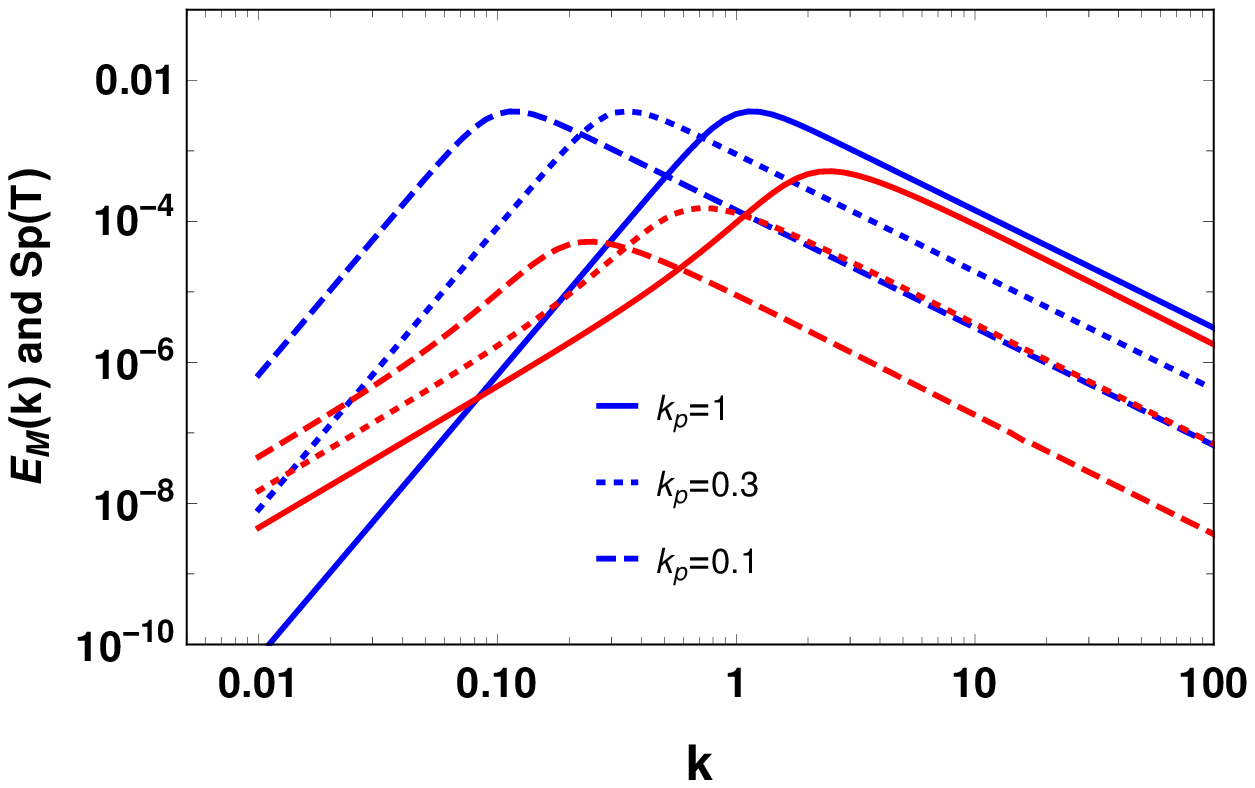}
\includegraphics[scale=0.7]{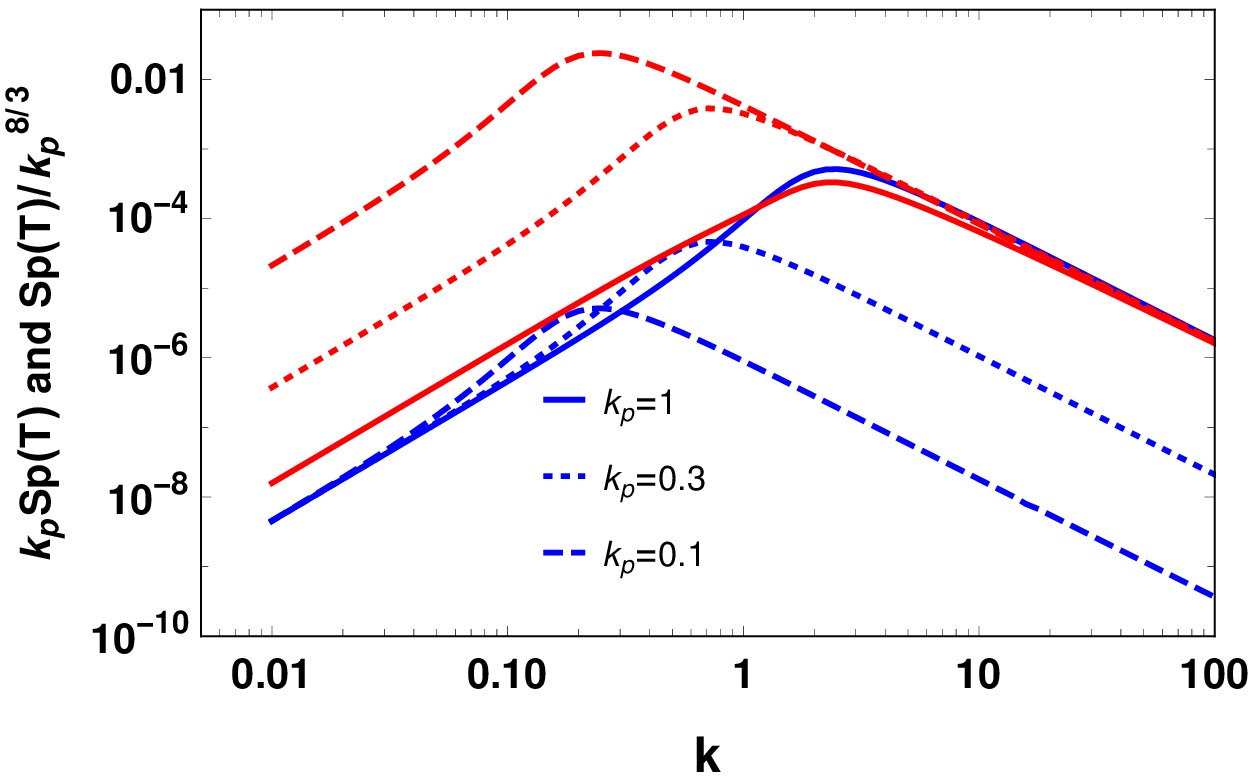}
\end{center}\caption{
Left: Solutions for $\Sp(\TT)$ (red) for different $\EM(k)$ (blue) for three
values of $k_{\rm p}$.
Right: solutions for $\Sp(\TT)$ scaled by $k_{\rm p}$ (blue) and
$k_{\rm p}^{-8/3}$ (red), to see its scalings in the subinertial
and inertial ranges, respectively.
}\label{pcascade}\end{figure*}

\begin{figure*}\begin{center}
\includegraphics[scale=0.7]{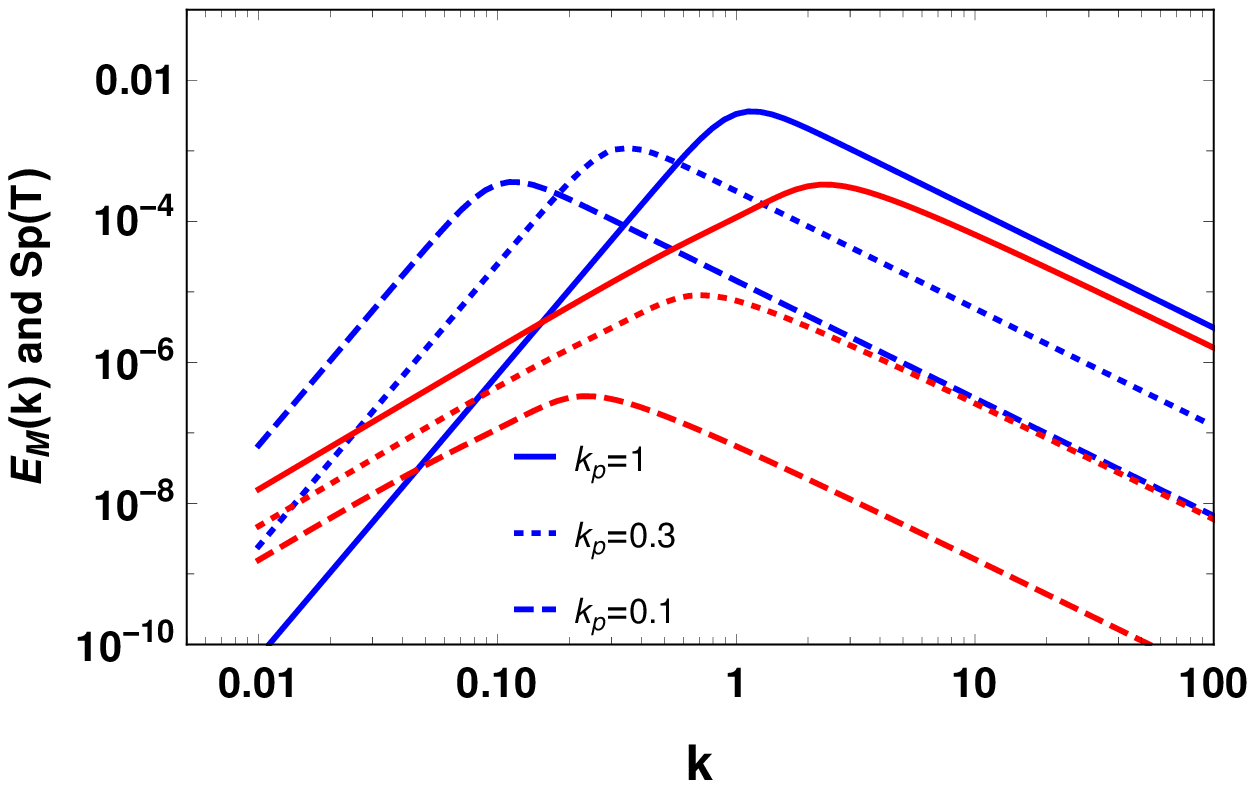}
\includegraphics[scale=0.7]{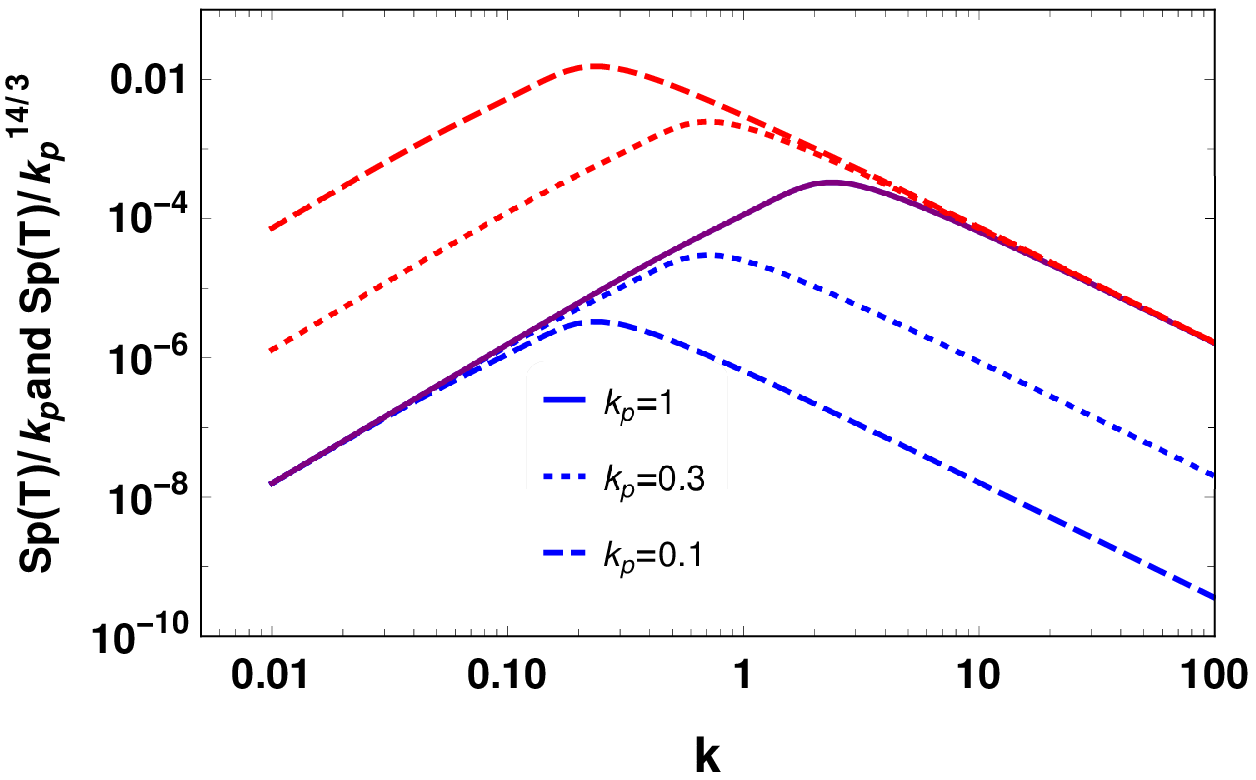}
\end{center}\caption{
Similar to \Fig{pcascade}, but for a case with $\beta=1$.
On the right, the solutions for $\Sp(\TT)$ are scaled by
$k_{\rm p}$ (blue) and $k_{\rm p}^{14/3}$ (red), to see its scalings
in the subinertial and inertial ranges, respectively.
Violet indicates the two overlap for $k_{\rm p}=1$.
}\label{pcascade_beta2}\end{figure*}

\subsection{Overall behavior of the stress}
\label{stress_evolution}

At the most minimalistic level, we can say that the magnetic field shows
an approximately self-similar evolution at late times,
where for $\runh$,
the peak value of $\EM(k,t)$ is unchanged, but the position
of the peak $k_{\rm p}$ goes to progressively smaller values as
$k_{\rm p}\sim t^{-2/3}$.

To understand the consequences for the evolution of the stress, let us now
consider an idealized model, where $\EM(k)\equiv\Sp(\BB)/2$ has a $k^4$
subinertial range, where $k<k_{\rm p}$, with $k_{\rm p}(t)$ being the
peak wave number, and a $k^{-5/3}$ inertial range spectrum for $k>k_{\rm p}$.
The spectrum of the transverse traceless part of the stress,
$\Sp(\TT)/2$, can be computed analytically using the expressions given
in \App{appendixa} (see Refs.~\cite{CDK04,Sharma+20} for details)
and is shown in \Figs{pcascade}{pcascade_beta2} for the helical and
nonhelical cases, respectively.
In these figures, we take three instances where the magnetic peaks are at
wave numbers $\kp=1$, 0.3, and $0.1$.

For HEL, the position of the peak of $\EM(k)$ is unchanged.
We see that, in agreement with earlier work \citep{RoperPol+20},
the positions of the peak of $\Sp(\TT)$ are always at $2k_{\rm p}$.
However, even though the peak values of $\EM(k)$ are unchanged, except
for the factor of two, those of $\Sp(\TT)$ are not and decay.
Nevertheless, at small $k$, $\Sp(\TT)$ still increases proportional to
$k_{\rm p}^{-1}$.
If $k_{\rm p}\propto t^{-2/3}$, as expected for helical turbulence
\citep{Hat84,BM99,BK17}, we find that $\Sp(\TT)\propto t^{2/3}$ for small $k$.

For NHEL, as shown in Ref.~\cite{BK17}, the peak of the
spectrum decreases with decreasing values of $k_{\rm p}$ proportional
to $k_{\rm p}^\beta$, where $\beta$ is an exponent that can be between
one and four.
In \Fig{pcascade_beta2}, we present the case with $\beta=1$ and find
that now $\Sp(\TT)(k)\propto k_{\rm p}$ for small $k$ and
$\propto k_{\rm p}^{14/3}$ for large $k$.
If $\kp\propto t^{-1/2}$, as expected for the nonhelical case for
$\beta=1$, $\Sp(\TT)\propto t^{-1/2}$ for small $k$.
We have summarized the behavior of $\Sp(\TT)$ with time in
\Tab{tildeT_vs_t} for helical and nonhelical cases.

\begin{table}\caption{
Time evolution of $\Sp(\TT)$ and $|\tilde{T}|$ from theory
}\begin{center}
\begin{tabular}{c|c|c}
&\multicolumn{1}{c}{helical}&
\multicolumn{1}{c}{nonhelical}\\
\hline
$\Sp(\TT)$ vs $k_{\rm p}$ at small $k$& $\propto 1/k_{\rm p}$& $\propto k_{\rm p}$ \\
\hline
$k_{\rm p}$ vs $t$ &$k_{\rm p} \propto t^{-2/3}$&$k_{\rm p} \propto t^{-1/2}$\\
\hline
$\Sp(\TT)$ vs $t$ & 
$\Sp(\TT) \propto t^{2/3}$& $\Sp(\TT)\propto t^{-1/2}$  \\
\hline
$|\tilde{T}|$ vs $t$ & 
$|\tilde{T}| \propto t^{1/3}$& $|\tilde{T}|\propto t^{-1/4}$ \\
\hline
\label{tildeT_vs_t}\end{tabular}
\end{center}
\end{table}

Recently, it has been found that the Hosking integral \cite{Sch20},
a Saffman-like helicity integral, is well
conserved in nonhelical magnetically dominated decaying turbulence
\citep{hosking20,ZBS22}, which implies $\beta=1.5$.
For the general case, we write $\Sp(\TT)\propto \kp^{2\beta-1}$
(for $k<\kp$), which implies $\Sp(\TT)\propto t^{-8/9}$ for
$\beta=1.5$ and $\kp\propto t^{-4/9}$.

It is also interesting to note that, for a given
spectrum of the magnetic field, $\Sp(\TT)$ is different for HEL and NHEL;
see the blue and red curves in \Fig{stress_spectrum_helvsnonhel}), respectively.
For the helical case, $\Sp(\TT)$ has smaller values compared to the nonhelical case
at wave numbers below $\kp$.
However, it has large values for wave numbers around the peak and above.
Such a feature of the stress spectrum also translates to the GW spectrum
and that is why we see a difference in the final GW spectrum produced
from helical and nonhelical cases discussed in the previous section.

\section{Predictions from algebraically growing stress}\label{simplemodel}

With the detailed information above, we are now in a position to compare
with the predictions from a simple time-dependent model.
In this section, we compute GW spectra by considering a simple
model for the time evolution of the stress.
It is assumed to increase algebraically as a power law
characterized by a power law index $\power$ during the time interval
from $t=1$ to $\te$.

\subsection{The model}

We model the $+$ and $\times$ polarizations of the Fourier-transformed
stress, $\tildeT(k,t)$, as
\begin{equation}\label{stresscase2}
    \tildeT(k,t)= \left\{\begin{array}{ll}
     |\tildeT_0(k)| \, t^{\power} , \quad & 1\le t \le \te, \\
    0, & t > \te,
    \end{array}\right.
\end{equation}
where $|\tildeT_0(k)|$
represents $|\tildeT(k,t)|$ at the initial time and is obtained
for given energy and helicity spectra of the magnetic field;
see \App{appendixa} for details.
We note that the authors of Ref.~\cite{RoperPol+22}
have developed an analytical model for the GW spectrum on the basis of
the time evolution of the stress, which they assumed constant during
a certain interval -- unlike our case.
The authors explain the location of certain breaks in their GW spectrum
as a consequence of the finite duration over which the stress is constant.
This duration is an empirical input parameter.
In our model, by contrast, the stress evolves as
a power law with an index that is in principle known from MHD theory,
although we can get even better agreement with the simulations when we
take the actual power-law index that is realized in the simulations.

\begin{figure*}
\includegraphics[scale=0.7]{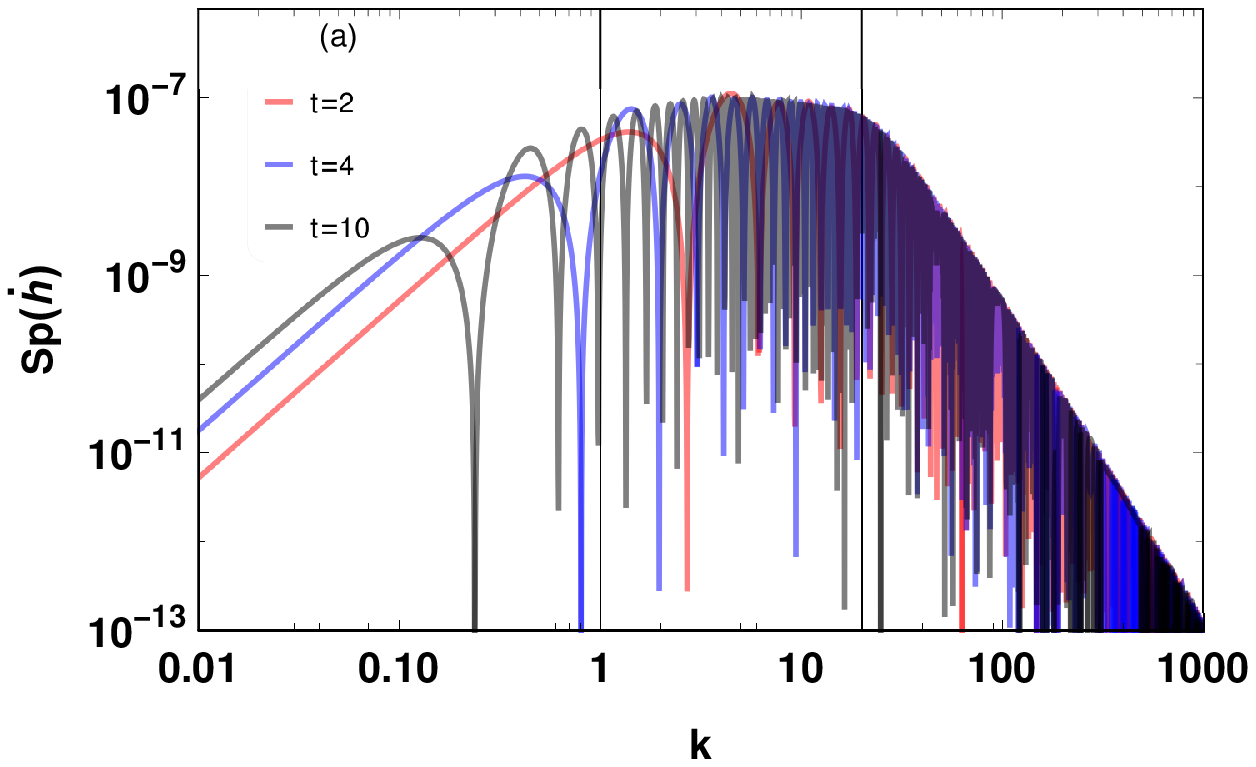}
\includegraphics[scale=0.7]{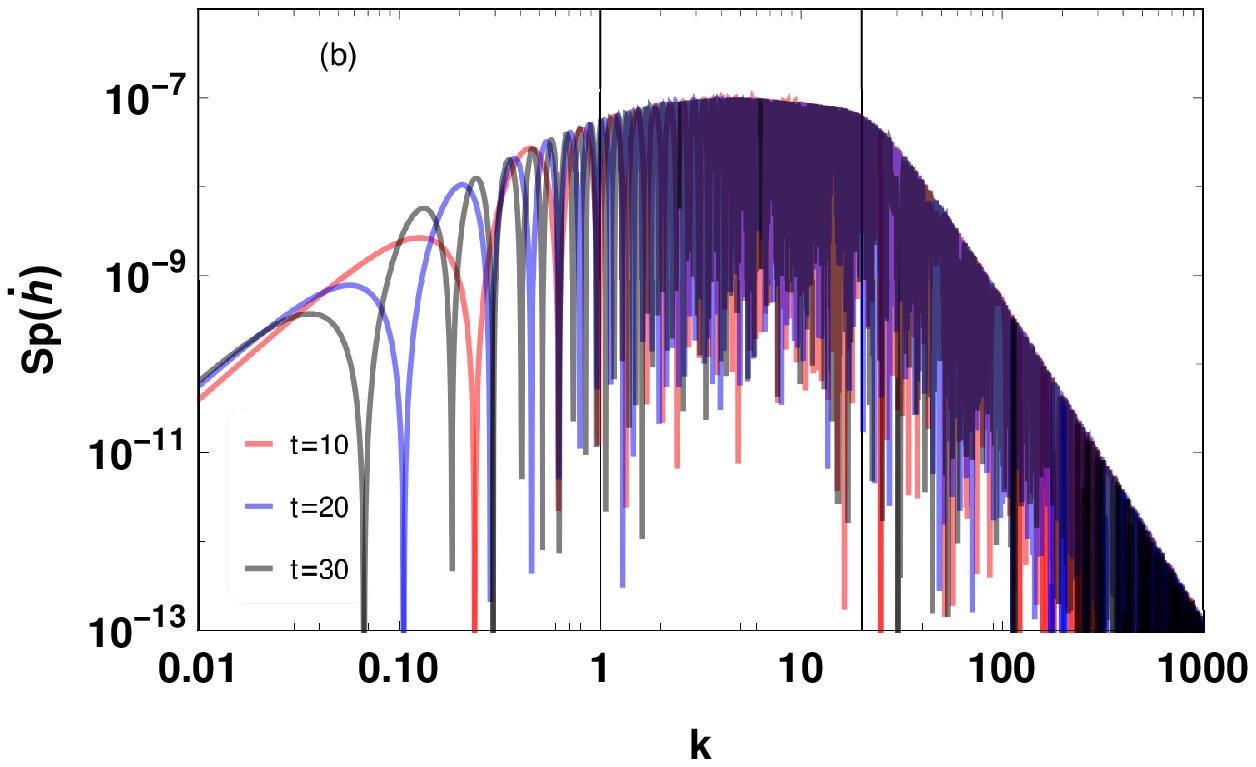}
\caption{
(a) $\Sp(\dot{\hh}(k,t)$ at different times.
Here, we assume $t=\te$,
$\EM=c (k/\kp)^4/(1+(k/\kp)^{17/3})$, where $\kp=10$
and $c=10^{-4}$ and $\power=-1/4$.
The red, blue, and black curves are for $\te=2$, $4$, and $10$ respectively.
The two black vertical lines correspond to $k_{*}$ and $2 \kp$.
(b) $\Sp(\dot{\hh})$ at times $\te=10$, $20$, and $30$.
}\label{results_from_model}
\end{figure*} 

To obtain the GW spectrum for our model, we first solve \Eq{GW4} for a
case when the source is active during the interval $1<t<\te$ and
thus obtain $\tildeh(k,t)$ and $\dot{\tildeh}(k,t)$.
The solution for $t\ge \te$ is given by
\begin{align}
    \tildeh(k,t)&=\int_1^{t} \frac{\sin k(t-t')}{k} \, \frac{6\tildeT(k,t')}{t'} \, dt', \\
    \dot{\tildeh}(k,t)&=\int_1^{t} \cos k(t-t') \, \frac{6\tildeT(k,t')}{t'} \, dt'. \label{hdot}
\end{align}
Using \Eq{hdot} and our model for $\tildeT(k,t)$, we obtain
\begin{align}
    \dot{\tildeh}(k,t)&=\frac{-3 |\tildeT_0(k)|}{(k t_0)^\power}\Big\{e^{i(kt-\power \pi/2)}\big[\Gamma(\power,ik\te)-\Gamma(\power,ikt_0)\big]\nonumber\\
    &+e^{-i(kt-\power \pi/2)}\big[\Gamma(\power,-ik\te)-\Gamma(\power,-ikt_0)\big]\Big\}.\label{full_solution}
\end{align}
In the above expression, $t_0=1$ represents the initial time.
In \Fig{results_from_model}(a), we show $\Sp(\dot{\hh})$
at different times for this model with $\power=-1/4$.
The red, blue, and black curves represent $\Sp(\dot{\hh})$ at $t=2$,
$4$, and $10$, respectively.
It is evident from this figure that $\Sp(\dot{\hh})$ is almost
flat for $1\la k\la 2 \kp$ and declines as $\propto k^{-11/3}$
for $k> 2\kp$.
$\Sp(\dot{\hh})$ is proportional to $k^2$ for $k<\kH$, where $\kH$
represents the wave number corresponding to the Hubble horizon size
at $t=\te$.
Further, as time increases, $\Sp(\dot{\hh})$ at low wave numbers
($\kH < k < 1$) grows and saturates, as is evident from
\Fig{results_from_model}(b).

\begin{figure}
\begin{center}
\includegraphics[scale=0.8]{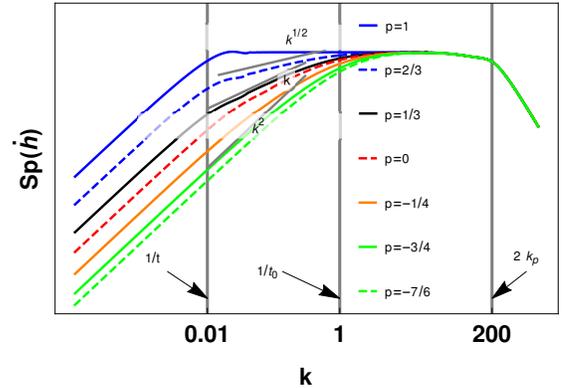}
\caption{$\Sp(\dot{\hh})$ for different values of $\power$. 
Here, the left, middle, and right black vertical lines represent the
wave numbers corresponding to the horizon size at the final time $t$,
the initial time $t_0=1$, and $2\kp$, respectively.}
\label{sphdot_for_different_gamma}
\end{center}
\end{figure}

To understand the role of the power-law index in the algebraically
growing part of the stress on $\Sp(\dot{\hh})$, we calculate
$\Sp(\dot{\hh})$ for different values of $\power$.
Those are shown in the right-hand panel of \Fig{sphdot_for_different_gamma}.
Here, $\Sp(\dot{\hh})$ is rapidly oscillating, so we plot
in this figure only its envelope.
From this figure, we conclude that $\Sp(\dot{\hh})$ can be divided
into three regime.
We begin discussing first the high wave number regime ($k>k_{0}$, regime~I),
where $k_{0}$ represents the wave number
corresponding to the Hubble horizon at the initial time. $\Sp(\dot{\hh})$
is flat and changes to $k^{-11/3}$ for $k>2\kp$.
For very low wave numbers corresponding to the superhorizon
range ($k<\kH$, regime~III) $\Sp(\dot{\hh})$ is proportional to $k^2$.
In the intermediate regime [$\kH\lesssim k\lesssim (1-\power)/t_0$, regime~II],
$\Sp(\dot{\hh})$ changes from a flat spectrum
to a $k^2$ spectrum as the wave number decreases.
Note that, as the wave numbers decrease, the transition from a flat spectrum
to a $k^2$ spectrum is faster for the case when $\power=-1/4$ than when it is $\power=1/3$. 
The wave number at which this transition occurs depends on the value of
$\power$ and can be understood as follows.
In the algebraically growing phase, the typical time scale over which
$\tildeT/t$ decays, is $\delta t_T\sim t/(1-\power)$ and the typical time
scale for sourcing GWs at a given wave number $k$
just after $t=1$ is $\delta t_{\rm GW}\sim 1/k$, as can be inferred
from the cosine function in \Eq{hdot}.
The value of $\tildeT/t$ does not change much when $\delta
t_{\rm GW}/\delta t_T\le 1$.
This implies that, for $k>(1-\power)/t_0$, there will be a finite
interval during which $\tildeT/t$ can be assumed constant.
However, for $k<(1-\power)/t_0$, $\tildeT/t$ always changes.
The wave number $k\sim (1-\power)/t_0$ corresponds to the wave number
where $\Sp(\dot{\hh})$ starts changing from a flat spectrum.

The nature of $\Sp(\dot{\hh})$ can also be understood
by writing the expression of $\dot{\tildeh}(k,t)$, given in
\Eq{full_solution}, for different limits depending on the values of $kt_0$
and $k\te$.
For $\te\gg t_0$, which is indeed the case, and $\power<1$,
\Eq{full_solution} reduces to
\begin{equation}
    \frac{\dot{\tildeh}(k,t)}{6 |\tildeT_0(k)|}\approx \left\{\begin{array}{lr}
     \frac{\sin{k(t-t_0)}}{kt_0} \quad \text{(I)},\\[5pt] 
  \frac{\Gamma[\power]}{(kt_0)^\power}\cos\left({kt-\frac{\power\pi}{2}}\right)-\frac{\cos{kt}}{\power} \quad \text{(II)},\\[5pt] 
   \frac{\cos{kt}}{\power}\left[\left(\frac{\te}{t_0}\right)^\power-1\right] \quad \text{(III)}. 
    \end{array}\right.
\end{equation}
Using this, we calculate the spectrum of $\dot{\tildeh}$; it is given by
\begin{equation}
   \frac{\Sp(\dot{\hh})(k,t)}{36 |\tildeT_0(k)|^2}\approx \left\{\begin{array}{lr}
    \left[\frac{\sin{k(t-t_0)}}{t_0}\right]^2 \quad \text{(I)}, \\[5pt]
  k^2\Big[-\frac{\cos{kt}}{\power}+\frac{\Gamma[\power]}{(kt_0)^\power}\cos({kt-\frac{\power\pi}{2}})\Big]^2 \, \text{(II)}, \\[5pt]
    k^2\Big\{\frac{\cos{kt}}{\power}\left[\left(\frac{\te}{t_0}\right)^\power-1\right]\Big\}^2 \quad \text{(III)}.
    \end{array}\right.\label{approx_solution}
\end{equation}
From the above expression, we conclude that the break points
for the different slopes of $\Sp(\dot{\hh})$
is decided by $|\tildeT_0(k)|^2$ for $k>t_0^{-1}$.
Here, $|\tildeT_0(k)|^2$ is flat for $k<2 \kp$ and
proportional to $k^{-11/3}$ for $k>2 \kp$.
For the superhorizon modes, i.e., $k<1/\te$, $\Sp(\dot{\hh})$
is proportional to $k^2$, and for wave numbers $t_0^{-1}<k<\te^{-1}$,
$\Sp(\dot{\hh})$ changes from a flat spectrum to $k^2$, as
shown as the blue curves in \Fig{sphdot_for_different_gamma}.

\begin{figure*}
\begin{center}
\includegraphics[scale=0.9]{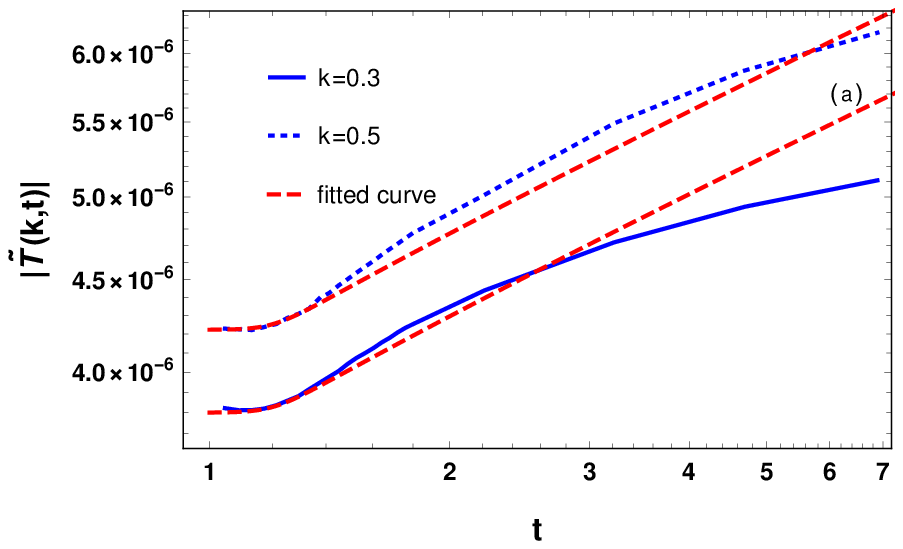}
\includegraphics[scale=0.9]{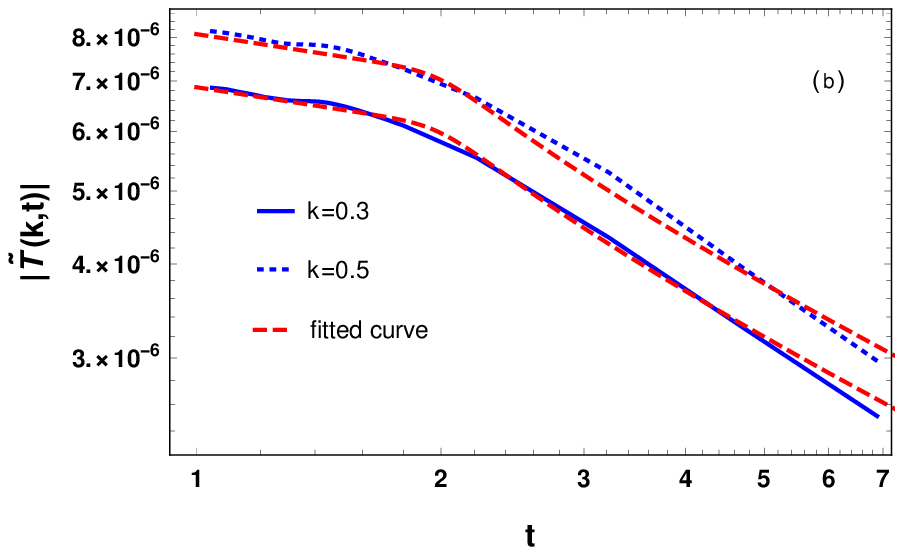}
\caption{
$\tildeT(k,t)$ vs.\ $t$. (a) 
The solid and dotted blue curves correspond to the time
evolution of $\tildeT(k,t)$ vs $t$ obtained from the simulation for 
$k=0.3$ and $0.5$, respectively,
for HEL and the red curve corresponds to a broken power law
fit to the blue curves.
(b) Same as (a), but for NHEL.
\label{t_evolution}
}\end{center}
\end{figure*}

\begin{figure*}
\begin{center}
\includegraphics[scale=0.9]{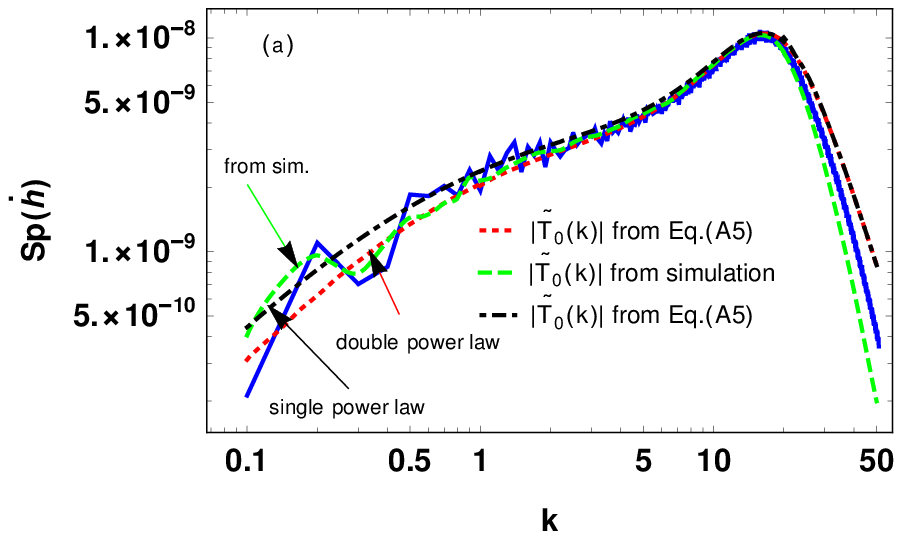}
\includegraphics[scale=0.9]{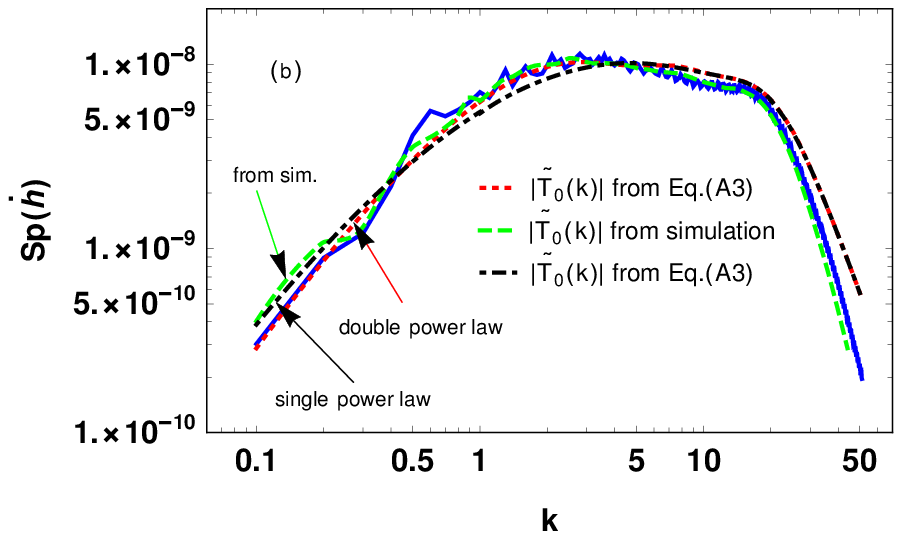}
\caption{
$\Sp(\dot{\hh})$ vs $k$: (a) The 
solid blue curve represents
$\Sp(\dot{\hh})$ corresponding to the run shown in
\Fig{pstress_etc_LowFreq_sig1}, respectively.
The dotted red and dashed green curves
represent the spectra obtained in the model
for the time evolution of $\tildeT(k,t)$ given in column 2 of
\Tab{table2}.
The dot dashed black curve represents $\Sp(\dot{\hh})$ for the time
evolution of $\tildeT(k,t)$ given in column 1 of \Tab{table2}.
The red curves are for the case when $|\tildeT_0(k)|$ is obtained using \Eq{tijh} of \App{appendixa}. For the green
curves, $|\tildeT_0(k)|$ is taken from the simulation.
(b) Same as (a), but for nonhelical case.
}\label{helical_comparison}
\end{center}
\end{figure*}

In this model, we take the same algebraic evolution with a constant
power-law index for all wave numbers.
In general, however, the time evolution of $\tildeT(k,t)$ is different for
wave numbers below and above the peak of $\Sp(\TT)(k,t)$.
For the case of helical magnetic fields discussed in \Fig{pcascade}, the
value of $\Sp(\TT)$ for a particular $k<2\kp$ at the initial time,
grows as $t^{2/3}$ until the time for which $\Sp(\TT)$ peaks at this
particular wave number.
After this time, the value of $\Sp(\TT)$ for this particular $k$
starts decreasing as $t^{-16/9}$.
This implies that we would have assumed $\tildeT(k)\propto t^{1/3}$
for $k<\kp$ and $\tildeT(k)\propto t^{-8/9}$ for $k>\kp$.
For the nonhelical magnetic field shown in \Fig{pcascade_beta2},
$\Sp(\TT)$ is always decreasing.
For $k<2 \kp$ at the
initial time, it first decreases as $t^{-1/2}$ and later switches to
$t^{-7/3}$.
We study $\Sp(\dot{\hh})$ by incorporating this and find that
there is no difference in the final $\Sp(\dot{\hh})$ compared to
the one obtained in our model.
This is why we did not consider the aforementioned evolution of the
stress to keep the model simple.

\section{Comparison of the analytical model with simulation results}
\label{comparison}

In the above section, we discussed $\Sp(\dot{\hh})$ in a model
inspired by the fact that the GW spectrum does not change if we change
the stress tensor by its modulus for decaying MHD turbulence in the
early Universe.
In this model, we approximate the stress tensor by
$|\tildeT(k,t)|\equiv\sqrt{\Sp(\TT)/4\pi k^2}$ and its time evolution
is parameterized as a power law with index $\power$; see \Fig{t_evolution}.
In this section, we provide a comparison for $\Sp(\dot{\hh})$
obtained in this model with the simulation results discussed in
\Sec{EvolutionOfStress}.
To compare, we show $\Sp(\dot{\hh})$ obtained from the
model and different simulations together.
In \Figsp{helical_comparison}{a}{b}, we plot the spectra for the runs
shown in \Figs{pstress_etc_LowFreq_sig1}{pstress_etc_LowFreq_sig0} and
discussed in \Sec{EvolutionOfStress}.
For the dotted-dashed black curve,
$|\tildeT_0(k)|$ is obtained by using \Eq{tijh} of \App{appendixa},
where we take $\EM=c (k/\kp)^4/[1+(k/\kp)^{17/3}]$ and the value of
the constant $c$ is determined such that the obtained stress spectrum
matches with that of the simulation at $t=1$.
For this case, the stress spectrum evolution is modeled as a single power
law and the value of $p$ is $1/3$ and $-1/4$ for the helical and nonhelical
cases, respectively.
The value of $p$ is decided from the time evolution of the low wave number
tail of $\Sp(\TT)$, as discussed in \Sec{stress_evolution}.
From this figure, it is concluded that the spectral nature of
$\Sp(\dot{\hh})$ matches well with the prediction from the model.
However, there is a difference at small wave numbers, especially for the
nonhelical case.
This is due to the fact that the modelling of $\Sp(\TT)$ by a single
power does not provide a better fit to the evolution obtained in $\runnh$.

A double power law of the form $t^{-1/3}/[1+(t-1)^n]^{5/7n}$, where
$n$ regulates the transition, here with $n=10$, provides a better fit
to $\Sp(\TT)$ for the low frequency tail for NHEL; see \Fig{t_evolution}.
In this figure, we plot $\tildeT(k,t)$ obtained from the simulation in solid and dotted blue for the wave numbers $k=0.3$ and $0.5$,
respectively and the double power law fit to the blue curves is in dashed
red color.
The double power law, which fits $\tildeT(k,t)$ for HEL,
is $1/[1+(t-0.2)^n]^{5/24n}$, where $n=20$.
After considering such a time evolution, the obtained $\Sp(\dot{\hh})$
is shown as the dotted red curve in \Fig{helical_comparison}(b).
For the dashed green curve, we consider $|\tildeT_0(k)|=\sqrt{\Sp(\TT)/4\pi k^2}$
and $\Sp(\TT)$ is obtained from the simulation at $t=1$.
These different forms of the time evolution of $\tildeT(k,t)$ are
given in \Tab{table2}. 
The spectra in \Figsp{helical_comparison}{a}{b} are plotted at a
time when the value of the $\Sp(\dot{\hh})$ for each wave number has
reached approximately a constant value.
The actual time for $\runh$ is $t=175.5$ and for $\runnh$ it is $t=99.5$.
For earlier times, the mean value of $\Sp(\dot{\hh})$ is in
reasonably good agreement with the values obtained from the model.

We notice that the nature of the GW spectra in $\runh$ and
$\runnh$ are different.
There is large power in $\runh$ compared to $\runnh$
around the peak of the GW spectrum for the same strength of the
initial magnetic field.
This is due to the presence of an additional term due to
the helicity spectrum in the stress spectrum; see \App{appendixa}.

\begin{table}\caption{
Time dependence of $\tildeT(k,t)$ taken in our analysis
}\begin{center}
\begin{tabular}{c|c|c}
run & from theory & from simulation\\
\hline
hel& $\left(\frac{t}{t_0}\right)^{1/3}$& $\frac{1}{(1+(t-0.2)^n)^{5/24n}}$ \\
\hline
nonhel& 
$\left(\frac{t}{t_0}\right)^{-1/4}$& $\frac{t^{-1/6}}{(1+(t-1)^n)^{5/14n}}$  \\
\hline
\label{table2}\end{tabular}
\end{center}
\end{table}

\section{Conclusions}\label{conclusion}

In this work, we have suggested a simple model to understand the GW spectrum
obtained for decaying MHD turbulence in the early Universe.
The Fourier-transformed stress is taken to be
$|\tilde{T}(k,t)|$, i.e., we ignore changes in the phase, and
its time evolution is parameterized by a power law.
Such a time evolution of the stress is motivated by the simulations for
the decaying MHD turbulence at low wave numbers discussed in \Sec{EvolutionOfStress}.
We find that the spectral nature of the GW spectrum is well represented
by this simple model.
In this work, we also show that the nature of the GW spectra in the helical
case are different from those in the nonhelical case.
Apart from the polarization of GW, this spectral difference may also
be important in distinguishing the helical and nonhelical nature of the
primordial magnetic field.

In this work, we have developed a model to understand the low frequency tail
of the GW spectrum in the cases where turbulence is initiated suddenly.
However, it will now also be interesting to study cases where the magnetic
field is generated selfconsistently, such as through the chiral magnetic
effect in the early Universe \citep{Roga_etal17,Schober+18}.
It would be interesting to see if a model such as the one discussed in
this paper can also explain the GW spectra obtained through the chiral
magnetic effect.
This, we hope to report in a future study.

\begin{acknowledgements}

RS would like to thank Hongzhe Zhou for helping him analyzing the output
data with the Mathematica routines of the {\sc Pencil Code}.
This work was supported by the Swedish Research Council
(Vetenskapsradet, 2019-04234).
Nordita is sponsored by Nordforsk.
We acknowledge the allocation of computing resources provided by the
Swedish National Allocations Committee at the Center for Parallel
Computers at the Royal Institute of Technology in Stockholm and
Link\"oping.

\end{acknowledgements}

\vspace{2mm}
{\bf Data availability}---The source code used for the
simulations of this study, the {\sc Pencil Code},
is freely available from Refs.~\cite{JOSS}.
Supplemental Material with plots similar to
\Figs{rslice_stress_plot_LowFreq}{rslice_stress_plot_LowFreq_nohel},
but for other wave vectors of the same length, along with the
simulation setups and the corresponding data
are freely available from Ref.~\cite{DATA}.

\appendix

\section{$\Sp(\TT)$ in terms of magnetic spectrum}
\label{appendixa}

Here, we provide the expressions for the stress spectrum in terms of the
magnetic spectrum for helical and nonhelical cases; see Refs.~\cite{CDK04}
and \cite{Sharma+20} for the derivation.
The two-point correlation for the nonhelical magnetic field in Fourier
space is given by,
\begin{align}
\langle \tilde{B}_{i}(\kk)\tilde{B}^*_{j}(\kk') \rangle &= (2 \pi)^3 \delta(\kk-\kk')(\delta_{ij}-\hat{\kk}_i \hat{\kk}_j)P_{\rm SM}(k)
\end{align}
Assuming a Gaussian nature of the magnetic field fluctuations,
the stress spectrum is given by
\begin{align}
\Sp(\TT)\equiv &4 \pi k^2\langle\tildeT_{ij}^{\rm TT}(\kk)\tildeT_{*ij}^{\rm TT}(\kk)\rangle=4\pi k^2\int d^3 q  \Big[ P_{\rm SM}(q)\nonumber\\
&P_{\rm SM}({|\kk-\qq|}) (1+\gamma^2+\beta^2+\gamma^2 \beta^2)\Big].
\end{align} 
Here, $\tildeT_{ij}^{\rm TT}(\kk)=-B_i B_j+\frac{1}{2}\delta_{ij}B_kB_k$.
In terms of the energy spectrum, $\EM(k)\equiv 4\pi k^2 P_{\rm SM}(k)$,
the above expression reduces to
\begin{align}\label{tijnh}
\Sp(\TT)&=\frac{1}{4\pi}\int d^3 q \frac{k^2}{q^2 |\kk-\qq|^2} \Big[ \EM(q)\nonumber\\
&\EM({|\kk-\qq|}) (1+\gamma^2+\beta^2+\gamma^2 \beta^2)\Big].
\end{align} 
In the above expressions $\gamma=\hat{\kk}\cdot \hat{\qq}$ and
$\beta=\hat{\kk}\cdot \widehat{\kk-\qq}$.
For helical magnetic fields, there is an additional antisymmetric
contribution to the two-point correlation and is given by,
\begin{align}
\langle \tilde{B}_{i}(\kk)\tilde{B}^*_{j}(\kk') \rangle&= (2 \pi)^3 \delta(\kk-\kk')\Big((\delta_{ij}-\hat{\kk}_i \hat{\kk}_j)P_{\rm SM}(k)\nonumber\\
&+i \epsilon_{ijm}\hat{\kk}_m P_{AM}(k) \Big).
\end{align}
The stress spectrum for this case is given by
\begin{align}
\Sp(\TT)=&4\pi k^2 \int d^3 q  \Big[ P_{\rm SM}(\qq)P_{\rm SM}({|\kk-\qq|})(1+\gamma^2+\beta^2+\gamma^2 \beta^2)\nonumber\\
&+4 \gamma \beta P_{\rm AM}(q)P_{\rm AM}({|\kk-\qq|})\Big].
\end{align} 
In terms of the energy spectrum, $\EM(k)$, and the helicity spectrum,
$H_M(k)\equiv 4\pi k^2 P_{\rm AM}(k)$, the above expression reduces to
\begin{align}\label{tijh}
\Sp(\TT)=&\frac{1}{4\pi}\int d^3 q \frac{k^2}{q^2 |\kk-\qq|^2}
\Big[ \EM(\qq)\EM({|\kk-\qq|})\nonumber\\&(1+\gamma^2+\beta^2+\gamma^2 \beta^2)+4 \gamma \beta H_M(q)H_M({|\kk-\qq|})\Big].
\end{align} 
\bibliography{ref}

\begin{thebibliography}{54}%
\makeatletter
\providecommand \@ifxundefined [1]{%
 \@ifx{#1\undefined}
}%
\providecommand \@ifnum [1]{%
 \ifnum #1\expandafter \@firstoftwo
 \else \expandafter \@secondoftwo
 \fi
}%
\providecommand \@ifx [1]{%
 \ifx #1\expandafter \@firstoftwo
 \else \expandafter \@secondoftwo
 \fi
}%
\providecommand \natexlab [1]{#1}%
\providecommand \enquote  [1]{``#1''}%
\providecommand \bibnamefont  [1]{#1}%
\providecommand \bibfnamefont [1]{#1}%
\providecommand \citenamefont [1]{#1}%
\providecommand \href@noop [0]{\@secondoftwo}%
\providecommand \href [0]{\begingroup \@sanitize@url \@href}%
\providecommand \@href[1]{\@@startlink{#1}\@@href}%
\providecommand \@@href[1]{\endgroup#1\@@endlink}%
\providecommand \@sanitize@url [0]{\catcode `\\12\catcode `\$12\catcode
  `\&12\catcode `\#12\catcode `\^12\catcode `\_12\catcode `\%12\relax}%
\providecommand \@@startlink[1]{}%
\providecommand \@@endlink[0]{}%
\providecommand \url  [0]{\begingroup\@sanitize@url \@url }%
\providecommand \@url [1]{\endgroup\@href {#1}{\urlprefix }}%
\providecommand \urlprefix  [0]{URL }%
\providecommand \Eprint [0]{\href }%
\providecommand \doibase [0]{http://dx.doi.org/}%
\providecommand \selectlanguage [0]{\@gobble}%
\providecommand \bibinfo  [0]{\@secondoftwo}%
\providecommand \bibfield  [0]{\@secondoftwo}%
\providecommand \translation [1]{[#1]}%
\providecommand \BibitemOpen [0]{}%
\providecommand \bibitemStop [0]{}%
\providecommand \bibitemNoStop [0]{.\EOS\space}%
\providecommand \EOS [0]{\spacefactor3000\relax}%
\providecommand \BibitemShut  [1]{\csname bibitem#1\endcsname}%
\let\auto@bib@innerbib\@empty
\bibitem [{\citenamefont {{Kamionkowski}}\ \emph {et~al.}(1994)\citenamefont
  {{Kamionkowski}}, \citenamefont {{Kosowsky}},\ and\ \citenamefont
  {{Turner}}}]{1994PhRvD..49.2837K}%
  \BibitemOpen
  \bibfield  {author} {\bibinfo {author} {\bibfnamefont {M.}~\bibnamefont
  {{Kamionkowski}}}, \bibinfo {author} {\bibfnamefont {A.}~\bibnamefont
  {{Kosowsky}}}, \ and\ \bibinfo {author} {\bibfnamefont {M.~S.}\ \bibnamefont
  {{Turner}}},\ }\href {\doibase 10.1103/PhysRevD.49.2837} {\bibfield
  {journal} {\bibinfo  {journal} {\prd}\ }\textbf {\bibinfo {volume} {49}},\
  \bibinfo {pages} {2837} (\bibinfo {year} {1994})},\ \Eprint
  {http://arxiv.org/abs/astro-ph/9310044} {arXiv:astro-ph/9310044 [astro-ph]}
  \BibitemShut {NoStop}%
\bibitem [{\citenamefont {Hogan}(2000)}]{PhysRevLett.85.2044}%
  \BibitemOpen
  \bibfield  {author} {\bibinfo {author} {\bibfnamefont {C.~J.}\ \bibnamefont
  {Hogan}},\ }\href {\doibase 10.1103/PhysRevLett.85.2044} {\bibfield
  {journal} {\bibinfo  {journal} {Phys. Rev. Lett.}\ }\textbf {\bibinfo
  {volume} {85}},\ \bibinfo {pages} {2044} (\bibinfo {year}
  {2000})}\BibitemShut {NoStop}%
\bibitem [{\citenamefont {{Kosowsky}}\ \emph {et~al.}(2002)\citenamefont
  {{Kosowsky}}, \citenamefont {{Mack}},\ and\ \citenamefont
  {{Kahniashvili}}}]{2002PhRvD..66b4030K}%
  \BibitemOpen
  \bibfield  {author} {\bibinfo {author} {\bibfnamefont {A.}~\bibnamefont
  {{Kosowsky}}}, \bibinfo {author} {\bibfnamefont {A.}~\bibnamefont {{Mack}}},
  \ and\ \bibinfo {author} {\bibfnamefont {T.}~\bibnamefont {{Kahniashvili}}},\
  }\href {\doibase 10.1103/PhysRevD.66.024030} {\bibfield  {journal} {\bibinfo
  {journal} {\prd}\ }\textbf {\bibinfo {volume} {66}},\ \bibinfo {eid} {024030}
  (\bibinfo {year} {2002})},\ \Eprint {http://arxiv.org/abs/astro-ph/0111483}
  {arXiv:astro-ph/0111483 [astro-ph]} \BibitemShut {NoStop}%
\bibitem [{\citenamefont {{Dolgov}}\ \emph {et~al.}(2002)\citenamefont
  {{Dolgov}}, \citenamefont {{Grasso}},\ and\ \citenamefont
  {{Nicolis}}}]{Dolgov+02}%
  \BibitemOpen
  \bibfield  {author} {\bibinfo {author} {\bibfnamefont {A.~D.}\ \bibnamefont
  {{Dolgov}}}, \bibinfo {author} {\bibfnamefont {D.}~\bibnamefont {{Grasso}}},
  \ and\ \bibinfo {author} {\bibfnamefont {A.}~\bibnamefont {{Nicolis}}},\
  }\href {\doibase 10.1103/PhysRevD.66.103505} {\bibfield  {journal} {\bibinfo
  {journal} {\prd}\ }\textbf {\bibinfo {volume} {66}},\ \bibinfo {eid} {103505}
  (\bibinfo {year} {2002})},\ \Eprint {http://arxiv.org/abs/astro-ph/0206461}
  {arXiv:astro-ph/0206461 [astro-ph]} \BibitemShut {NoStop}%
\bibitem [{\citenamefont {{Caprini}}\ and\ \citenamefont
  {{Figueroa}}(2018)}]{2018CQGra..35p3001C}%
  \BibitemOpen
  \bibfield  {author} {\bibinfo {author} {\bibfnamefont {C.}~\bibnamefont
  {{Caprini}}}\ and\ \bibinfo {author} {\bibfnamefont {D.~G.}\ \bibnamefont
  {{Figueroa}}},\ }\href {\doibase 10.1088/1361-6382/aac608} {\bibfield
  {journal} {\bibinfo  {journal} {CQGra}\ }\textbf {\bibinfo {volume} {35}},\
  \bibinfo {eid} {163001} (\bibinfo {year} {2018})},\ \Eprint
  {http://arxiv.org/abs/1801.04268} {arXiv:1801.04268 [astro-ph.CO]}
  \BibitemShut {NoStop}%
\bibitem [{\citenamefont {Kajantie}\ \emph {et~al.}(1996)\citenamefont
  {Kajantie}, \citenamefont {Laine}, \citenamefont {Rummukainen},\ and\
  \citenamefont {Shaposhnikov}}]{PhysRevLett.77.2887}%
  \BibitemOpen
  \bibfield  {author} {\bibinfo {author} {\bibfnamefont {K.}~\bibnamefont
  {Kajantie}}, \bibinfo {author} {\bibfnamefont {M.}~\bibnamefont {Laine}},
  \bibinfo {author} {\bibfnamefont {K.}~\bibnamefont {Rummukainen}}, \ and\
  \bibinfo {author} {\bibfnamefont {M.}~\bibnamefont {Shaposhnikov}},\ }\href
  {\doibase 10.1103/PhysRevLett.77.2887} {\bibfield  {journal} {\bibinfo
  {journal} {Phys. Rev. Lett.}\ }\textbf {\bibinfo {volume} {77}},\ \bibinfo
  {pages} {2887} (\bibinfo {year} {1996})}\BibitemShut {NoStop}%
\bibitem [{\citenamefont {{Aoki}}\ \emph {et~al.}(2006)\citenamefont {{Aoki}},
  \citenamefont {{Endr{\H{o}}di}}, \citenamefont {{Fodor}}, \citenamefont
  {{Katz}},\ and\ \citenamefont {{Szab{\'o}}}}]{2006Natur.443..675A}%
  \BibitemOpen
  \bibfield  {author} {\bibinfo {author} {\bibfnamefont {Y.}~\bibnamefont
  {{Aoki}}}, \bibinfo {author} {\bibfnamefont {G.}~\bibnamefont
  {{Endr{\H{o}}di}}}, \bibinfo {author} {\bibfnamefont {Z.}~\bibnamefont
  {{Fodor}}}, \bibinfo {author} {\bibfnamefont {S.~D.}\ \bibnamefont {{Katz}}},
  \ and\ \bibinfo {author} {\bibfnamefont {K.~K.}\ \bibnamefont
  {{Szab{\'o}}}},\ }\href {\doibase 10.1038/nature05120} {\bibfield  {journal}
  {\bibinfo  {journal} {\nat}\ }\textbf {\bibinfo {volume} {443}},\ \bibinfo
  {pages} {675} (\bibinfo {year} {2006})},\ \Eprint
  {http://arxiv.org/abs/hep-lat/0611014} {arXiv:hep-lat/0611014 [hep-lat]}
  \BibitemShut {NoStop}%
\bibitem [{\citenamefont {Witten}(1984)}]{PhysRevD.30.272}%
  \BibitemOpen
  \bibfield  {author} {\bibinfo {author} {\bibfnamefont {E.}~\bibnamefont
  {Witten}},\ }\href {\doibase 10.1103/PhysRevD.30.272} {\bibfield  {journal}
  {\bibinfo  {journal} {Phys. Rev. D}\ }\textbf {\bibinfo {volume} {30}},\
  \bibinfo {pages} {272} (\bibinfo {year} {1984})}\BibitemShut {NoStop}%
\bibitem [{\citenamefont {{Hogan}}(1986)}]{10.1093/mnras/218.4.629}%
  \BibitemOpen
  \bibfield  {author} {\bibinfo {author} {\bibfnamefont {C.~J.}\ \bibnamefont
  {{Hogan}}},\ }\href {\doibase 10.1093/mnras/218.4.629} {\bibfield  {journal}
  {\bibinfo  {journal} {\mnras}\ }\textbf {\bibinfo {volume} {218}},\ \bibinfo
  {pages} {629} (\bibinfo {year} {1986})}\BibitemShut {NoStop}%
\bibitem [{\citenamefont {Mazumdar}\ and\ \citenamefont
  {White}(2019)}]{Mazumdar_2019}%
  \BibitemOpen
  \bibfield  {author} {\bibinfo {author} {\bibfnamefont {A.}~\bibnamefont
  {Mazumdar}}\ and\ \bibinfo {author} {\bibfnamefont {G.}~\bibnamefont
  {White}},\ }\href {\doibase 10.1088/1361-6633/ab1f55} {\bibfield  {journal}
  {\bibinfo  {journal} {Rep. Prog. Phys.}\ }\textbf {\bibinfo {volume} {82}},\
  \bibinfo {pages} {076901} (\bibinfo {year} {2019})}\BibitemShut {NoStop}%
\bibitem [{\citenamefont {Hindmarsh}\ \emph {et~al.}(2021)\citenamefont
  {Hindmarsh}, \citenamefont {Lüben}, \citenamefont {Lumma},\ and\
  \citenamefont {Pauly}}]{10.21468/SciPostPhysLectNotes.24}%
  \BibitemOpen
  \bibfield  {author} {\bibinfo {author} {\bibfnamefont {M.}~\bibnamefont
  {Hindmarsh}}, \bibinfo {author} {\bibfnamefont {M.}~\bibnamefont {Lüben}},
  \bibinfo {author} {\bibfnamefont {J.}~\bibnamefont {Lumma}}, \ and\ \bibinfo
  {author} {\bibfnamefont {M.}~\bibnamefont {Pauly}},\ }\href {\doibase
  10.21468/SciPostPhysLectNotes.24} {\bibfield  {journal} {\bibinfo  {journal}
  {SciPost Phys. Lect. Notes}\ ,\ \bibinfo {pages} {24}} (\bibinfo {year}
  {2021})}\BibitemShut {NoStop}%
\bibitem [{\citenamefont {{Turner}}\ and\ \citenamefont
  {{Widrow}}(1988)}]{1988PhRvD..37.2743T}%
  \BibitemOpen
  \bibfield  {author} {\bibinfo {author} {\bibfnamefont {M.~S.}\ \bibnamefont
  {{Turner}}}\ and\ \bibinfo {author} {\bibfnamefont {L.~M.}\ \bibnamefont
  {{Widrow}}},\ }\href {\doibase 10.1103/PhysRevD.37.2743} {\bibfield
  {journal} {\bibinfo  {journal} {\prd}\ }\textbf {\bibinfo {volume} {37}},\
  \bibinfo {pages} {2743} (\bibinfo {year} {1988})}\BibitemShut {NoStop}%
\bibitem [{\citenamefont {{Vachaspati}}(1991)}]{tanmay1991}%
  \BibitemOpen
  \bibfield  {author} {\bibinfo {author} {\bibfnamefont {T.}~\bibnamefont
  {{Vachaspati}}},\ }\href {\doibase 10.1016/0370-2693(91)90051-Q} {\bibfield
  {journal} {\bibinfo  {journal} {PhLB}\ }\textbf {\bibinfo {volume} {265}},\
  \bibinfo {pages} {258} (\bibinfo {year} {1991})}\BibitemShut {NoStop}%
\bibitem [{\citenamefont {{Ratra}}(1992)}]{1992ApJ...391L...1R}%
  \BibitemOpen
  \bibfield  {author} {\bibinfo {author} {\bibfnamefont {B.}~\bibnamefont
  {{Ratra}}},\ }\href {\doibase 10.1086/186384} {\bibfield  {journal} {\bibinfo
   {journal} {\apjl}\ }\textbf {\bibinfo {volume} {391}},\ \bibinfo {pages}
  {L1} (\bibinfo {year} {1992})}\BibitemShut {NoStop}%
\bibitem [{\citenamefont {Durrer}\ and\ \citenamefont
  {Neronov}(2013)}]{durrer2013}%
  \BibitemOpen
  \bibfield  {author} {\bibinfo {author} {\bibfnamefont {R.}~\bibnamefont
  {Durrer}}\ and\ \bibinfo {author} {\bibfnamefont {A.}~\bibnamefont
  {Neronov}},\ }\href {\doibase 10.1007/s00159-013-0062-7} {\bibfield
  {journal} {\bibinfo  {journal} {Astron. Astrophys. Rev.}\ }\textbf {\bibinfo
  {volume} {21}},\ \bibinfo {pages} {62} (\bibinfo {year} {2013})},\ \Eprint
  {http://arxiv.org/abs/1303.7121} {arXiv:1303.7121 [astro-ph.CO]} \BibitemShut
  {NoStop}%
\bibitem [{\citenamefont {Subramanian}(2016)}]{subramanian2016}%
  \BibitemOpen
  \bibfield  {author} {\bibinfo {author} {\bibfnamefont {K.}~\bibnamefont
  {Subramanian}},\ }\href {\doibase 10.1088/0034-4885/79/7/076901} {\bibfield
  {journal} {\bibinfo  {journal} {Rep. Prog. Phys.}\ }\textbf {\bibinfo
  {volume} {79}},\ \bibinfo {pages} {076901} (\bibinfo {year}
  {2016})}\BibitemShut {NoStop}%
\bibitem [{\citenamefont {{Subramanian}}(2019)}]{kandu2019}%
  \BibitemOpen
  \bibfield  {author} {\bibinfo {author} {\bibfnamefont {K.}~\bibnamefont
  {{Subramanian}}},\ }\href {\doibase 10.3390/galaxies7020047} {\bibfield
  {journal} {\bibinfo  {journal} {Galaxies}\ }\textbf {\bibinfo {volume} {7}},\
  \bibinfo {pages} {47} (\bibinfo {year} {2019})},\ \Eprint
  {http://arxiv.org/abs/1903.03744} {arXiv:1903.03744 [astro-ph.CO]}
  \BibitemShut {NoStop}%
\bibitem [{\citenamefont {{Vachaspati}}(2021)}]{tanmay2021}%
  \BibitemOpen
  \bibfield  {author} {\bibinfo {author} {\bibfnamefont {T.}~\bibnamefont
  {{Vachaspati}}},\ }\href {\doibase 10.1088/1361-6633/ac03a9} {\bibfield
  {journal} {\bibinfo  {journal} {Reports on Progress in Physics}\ }\textbf
  {\bibinfo {volume} {84}},\ \bibinfo {eid} {074901} (\bibinfo {year}
  {2021})},\ \Eprint {http://arxiv.org/abs/2010.10525} {arXiv:2010.10525
  [astro-ph.CO]} \BibitemShut {NoStop}%
\bibitem [{\citenamefont {{Brandenburg}}\ \emph {et~al.}(2021)\citenamefont
  {{Brandenburg}}, \citenamefont {{He}},\ and\ \citenamefont
  {{Sharma}}}]{Bran+He+Shar21}%
  \BibitemOpen
  \bibfield  {author} {\bibinfo {author} {\bibfnamefont {A.}~\bibnamefont
  {{Brandenburg}}}, \bibinfo {author} {\bibfnamefont {Y.}~\bibnamefont {{He}}},
  \ and\ \bibinfo {author} {\bibfnamefont {R.}~\bibnamefont {{Sharma}}},\
  }\href {\doibase 10.3847/1538-4357/ac20d9} {\bibfield  {journal} {\bibinfo
  {journal} {\apj}\ }\textbf {\bibinfo {volume} {922}},\ \bibinfo {eid} {192}
  (\bibinfo {year} {2021})},\ \Eprint {http://arxiv.org/abs/2107.12333}
  {arXiv:2107.12333 [astro-ph.CO]} \BibitemShut {NoStop}%
\bibitem [{\citenamefont {Arzoumanian}\ \emph {et~al.}(2020)\citenamefont
  {Arzoumanian} \emph {et~al.}}]{NANOGrav2020}%
  \BibitemOpen
  \bibfield  {author} {\bibinfo {author} {\bibfnamefont {Z.}~\bibnamefont
  {Arzoumanian}} \emph {et~al.} (\bibinfo {collaboration} {NANOGrav}),\ }\href
  {\doibase 10.3847/2041-8213/abd401} {\bibfield  {journal} {\bibinfo
  {journal} {Astrophys. J. Lett.}\ }\textbf {\bibinfo {volume} {905}},\
  \bibinfo {pages} {L34} (\bibinfo {year} {2020})},\ \Eprint
  {http://arxiv.org/abs/2009.04496} {arXiv:2009.04496 [astro-ph.HE]}
  \BibitemShut {NoStop}%
\bibitem [{\citenamefont {Goncharov}\ \emph {et~al.}(2021)\citenamefont
  {Goncharov} \emph {et~al.}}]{Goncharov_2021}%
  \BibitemOpen
  \bibfield  {author} {\bibinfo {author} {\bibfnamefont {B.}~\bibnamefont
  {Goncharov}} \emph {et~al.},\ }\href {\doibase 10.3847/2041-8213/ac17f4}
  {\bibfield  {journal} {\bibinfo  {journal} {Astrophys. J. Lett.}\ }\textbf
  {\bibinfo {volume} {917}},\ \bibinfo {pages} {L19} (\bibinfo {year}
  {2021})},\ \Eprint {http://arxiv.org/abs/2107.12112} {arXiv:2107.12112
  [astro-ph.HE]} \BibitemShut {NoStop}%
\bibitem [{\citenamefont {Chen}\ \emph {et~al.}(2021)\citenamefont {Chen} \emph
  {et~al.}}]{Chen2021}%
  \BibitemOpen
  \bibfield  {author} {\bibinfo {author} {\bibfnamefont {S.}~\bibnamefont
  {Chen}} \emph {et~al.},\ }\href {\doibase 10.1093/mnras/stab2833} {\bibfield
  {journal} {\bibinfo  {journal} {Mon. Not. Roy. Astron. Soc.}\ }\textbf
  {\bibinfo {volume} {508}},\ \bibinfo {pages} {4970} (\bibinfo {year}
  {2021})},\ \Eprint {http://arxiv.org/abs/2110.13184} {arXiv:2110.13184
  [astro-ph.HE]} \BibitemShut {NoStop}%
\bibitem [{\citenamefont {Antoniadis}\ \emph {et~al.}(2022)\citenamefont
  {Antoniadis} \emph {et~al.}}]{Antoniadis2022}%
  \BibitemOpen
  \bibfield  {author} {\bibinfo {author} {\bibfnamefont {J.}~\bibnamefont
  {Antoniadis}} \emph {et~al.},\ }\href {\doibase 10.1093/mnras/stab3418}
  {\bibfield  {journal} {\bibinfo  {journal} {Mon. Not. Roy. Astron. Soc.}\
  }\textbf {\bibinfo {volume} {510}},\ \bibinfo {pages} {4873} (\bibinfo {year}
  {2022})},\ \Eprint {http://arxiv.org/abs/2201.03980} {arXiv:2201.03980
  [astro-ph.HE]} \BibitemShut {NoStop}%
\bibitem [{\citenamefont {Neronov}\ \emph {et~al.}(2021)\citenamefont
  {Neronov}, \citenamefont {Roper~Pol}, \citenamefont {Caprini},\ and\
  \citenamefont {Semikoz}}]{2021PhRvD.103L1302N}%
  \BibitemOpen
  \bibfield  {author} {\bibinfo {author} {\bibfnamefont {A.}~\bibnamefont
  {Neronov}}, \bibinfo {author} {\bibfnamefont {A.}~\bibnamefont {Roper~Pol}},
  \bibinfo {author} {\bibfnamefont {C.}~\bibnamefont {Caprini}}, \ and\
  \bibinfo {author} {\bibfnamefont {D.}~\bibnamefont {Semikoz}},\ }\href
  {\doibase 10.1103/PhysRevD.103.L041302} {\bibfield  {journal} {\bibinfo
  {journal} {Phys. Rev. D}\ }\textbf {\bibinfo {volume} {103}},\ \bibinfo
  {pages} {L041302} (\bibinfo {year} {2021})}\BibitemShut {NoStop}%
\bibitem [{\citenamefont {{Sharma}}(2022)}]{Sharma21}%
  \BibitemOpen
  \bibfield  {author} {\bibinfo {author} {\bibfnamefont {R.}~\bibnamefont
  {{Sharma}}},\ }\href {\doibase 10.1103/PhysRevD.105.L041302} {\bibfield
  {journal} {\bibinfo  {journal} {\prd}\ }\textbf {\bibinfo {volume} {105}},\
  \bibinfo {eid} {L041302} (\bibinfo {year} {2022})},\ \Eprint
  {http://arxiv.org/abs/2102.09358} {arXiv:2102.09358 [astro-ph.CO]}
  \BibitemShut {NoStop}%
\bibitem [{\citenamefont {{Roper Pol}}\ \emph
  {et~al.}(2022{\natexlab{a}})\citenamefont {{Roper Pol}}, \citenamefont
  {{Caprini}}, \citenamefont {{Neronov}},\ and\ \citenamefont
  {{Semikoz}}}]{RoperPol+22}%
  \BibitemOpen
  \bibfield  {author} {\bibinfo {author} {\bibfnamefont {A.}~\bibnamefont
  {{Roper Pol}}}, \bibinfo {author} {\bibfnamefont {C.}~\bibnamefont
  {{Caprini}}}, \bibinfo {author} {\bibfnamefont {A.}~\bibnamefont
  {{Neronov}}}, \ and\ \bibinfo {author} {\bibfnamefont {D.}~\bibnamefont
  {{Semikoz}}},\ }\href {\doibase 10.1103/PhysRevD.105.123502} {\bibfield
  {journal} {\bibinfo  {journal} {\prd}\ }\textbf {\bibinfo {volume} {105}},\
  \bibinfo {eid} {123502} (\bibinfo {year} {2022}{\natexlab{a}})},\ \Eprint
  {http://arxiv.org/abs/2201.05630} {arXiv:2201.05630 [astro-ph.CO]}
  \BibitemShut {NoStop}%
\bibitem [{\citenamefont {{Hellings}}\ and\ \citenamefont
  {{Downs}}(1983)}]{Hellings&Downs}%
  \BibitemOpen
  \bibfield  {author} {\bibinfo {author} {\bibfnamefont {R.~W.}\ \bibnamefont
  {{Hellings}}}\ and\ \bibinfo {author} {\bibfnamefont {G.~S.}\ \bibnamefont
  {{Downs}}},\ }\href {\doibase 10.1086/183954} {\bibfield  {journal} {\bibinfo
   {journal} {\apjl}\ }\textbf {\bibinfo {volume} {265}},\ \bibinfo {pages}
  {L39} (\bibinfo {year} {1983})}\BibitemShut {NoStop}%
\bibitem [{\citenamefont {{Roper Pol}}\ \emph
  {et~al.}(2020{\natexlab{a}})\citenamefont {{Roper Pol}}, \citenamefont
  {{Mandal}}, \citenamefont {{Brandenburg}}, \citenamefont {{Kahniashvili}},\
  and\ \citenamefont {{Kosowsky}}}]{RoperPol+20}%
  \BibitemOpen
  \bibfield  {author} {\bibinfo {author} {\bibfnamefont {A.}~\bibnamefont
  {{Roper Pol}}}, \bibinfo {author} {\bibfnamefont {S.}~\bibnamefont
  {{Mandal}}}, \bibinfo {author} {\bibfnamefont {A.}~\bibnamefont
  {{Brandenburg}}}, \bibinfo {author} {\bibfnamefont {T.}~\bibnamefont
  {{Kahniashvili}}}, \ and\ \bibinfo {author} {\bibfnamefont {A.}~\bibnamefont
  {{Kosowsky}}},\ }\href {\doibase 10.1103/PhysRevD.102.083512} {\bibfield
  {journal} {\bibinfo  {journal} {\prd}\ }\textbf {\bibinfo {volume} {102}},\
  \bibinfo {eid} {083512} (\bibinfo {year} {2020}{\natexlab{a}})},\ \Eprint
  {http://arxiv.org/abs/1903.08585} {arXiv:1903.08585 [astro-ph.CO]}
  \BibitemShut {NoStop}%
\bibitem [{\citenamefont {Auclair}\ \emph {et~al.}(2022)\citenamefont
  {Auclair}, \citenamefont {Caprini}, \citenamefont {Cutting}, \citenamefont
  {Hindmarsh}, \citenamefont {Rummukainen}, \citenamefont {Steer},\ and\
  \citenamefont {Weir}}]{Auclair:2022jod}%
  \BibitemOpen
  \bibfield  {author} {\bibinfo {author} {\bibfnamefont {P.}~\bibnamefont
  {Auclair}}, \bibinfo {author} {\bibfnamefont {C.}~\bibnamefont {Caprini}},
  \bibinfo {author} {\bibfnamefont {D.}~\bibnamefont {Cutting}}, \bibinfo
  {author} {\bibfnamefont {M.}~\bibnamefont {Hindmarsh}}, \bibinfo {author}
  {\bibfnamefont {K.}~\bibnamefont {Rummukainen}}, \bibinfo {author}
  {\bibfnamefont {D.~A.}\ \bibnamefont {Steer}}, \ and\ \bibinfo {author}
  {\bibfnamefont {D.~J.}\ \bibnamefont {Weir}},\ }\href@noop {} {\  (\bibinfo
  {year} {2022})},\ \Eprint {http://arxiv.org/abs/2205.02588} {arXiv:2205.02588
  [astro-ph.CO]} \BibitemShut {NoStop}%
\bibitem [{\citenamefont {{Brandenburg}}\ \emph {et~al.}(2015)\citenamefont
  {{Brandenburg}}, \citenamefont {{Kahniashvili}},\ and\ \citenamefont
  {{Tevzadze}}}]{2015PhRvL.114g5001B}%
  \BibitemOpen
  \bibfield  {author} {\bibinfo {author} {\bibfnamefont {A.}~\bibnamefont
  {{Brandenburg}}}, \bibinfo {author} {\bibfnamefont {T.}~\bibnamefont
  {{Kahniashvili}}}, \ and\ \bibinfo {author} {\bibfnamefont {A.~G.}\
  \bibnamefont {{Tevzadze}}},\ }\href {\doibase 10.1103/PhysRevLett.114.075001}
  {\bibfield  {journal} {\bibinfo  {journal} {PhRvL}\ }\textbf {\bibinfo
  {volume} {114}},\ \bibinfo {eid} {075001} (\bibinfo {year} {2015})},\ \Eprint
  {http://arxiv.org/abs/1404.2238} {arXiv:1404.2238} \BibitemShut {NoStop}%
\bibitem [{\citenamefont {{Zrake}}(2014)}]{2014ApJ...794L..26Z}%
  \BibitemOpen
  \bibfield  {author} {\bibinfo {author} {\bibfnamefont {J.}~\bibnamefont
  {{Zrake}}},\ }\href {\doibase 10.1088/2041-8205/794/2/L26} {\bibfield
  {journal} {\bibinfo  {journal} {\apjl}\ }\textbf {\bibinfo {volume} {794}},\
  \bibinfo {eid} {L26} (\bibinfo {year} {2014})},\ \Eprint
  {http://arxiv.org/abs/1407.5626} {arXiv:1407.5626 [astro-ph.HE]} \BibitemShut
  {NoStop}%
\bibitem [{\citenamefont {Kosowsky}\ \emph {et~al.}(2002)\citenamefont
  {Kosowsky}, \citenamefont {Mack},\ and\ \citenamefont
  {Kahniashvili}}]{kosowsky2002}%
  \BibitemOpen
  \bibfield  {author} {\bibinfo {author} {\bibfnamefont {A.}~\bibnamefont
  {Kosowsky}}, \bibinfo {author} {\bibfnamefont {A.}~\bibnamefont {Mack}}, \
  and\ \bibinfo {author} {\bibfnamefont {T.}~\bibnamefont {Kahniashvili}},\
  }\href {\doibase 10.1103/PhysRevD.66.024030} {\bibfield  {journal} {\bibinfo
  {journal} {PhRvD}\ }\textbf {\bibinfo {volume} {66}},\ \bibinfo {pages}
  {024030} (\bibinfo {year} {2002})}\BibitemShut {NoStop}%
\bibitem [{\citenamefont {{Gogoberidze}}\ \emph {et~al.}(2007)\citenamefont
  {{Gogoberidze}}, \citenamefont {{Kahniashvili}},\ and\ \citenamefont
  {{Kosowsky}}}]{Gogo+07}%
  \BibitemOpen
  \bibfield  {author} {\bibinfo {author} {\bibfnamefont {G.}~\bibnamefont
  {{Gogoberidze}}}, \bibinfo {author} {\bibfnamefont {T.}~\bibnamefont
  {{Kahniashvili}}}, \ and\ \bibinfo {author} {\bibfnamefont {A.}~\bibnamefont
  {{Kosowsky}}},\ }\href {\doibase 10.1103/PhysRevD.76.083002} {\bibfield
  {journal} {\bibinfo  {journal} {\prd}\ }\textbf {\bibinfo {volume} {76}},\
  \bibinfo {eid} {083002} (\bibinfo {year} {2007})},\ \Eprint
  {http://arxiv.org/abs/0705.1733} {arXiv:0705.1733 [astro-ph]} \BibitemShut
  {NoStop}%
\bibitem [{\citenamefont {Kahniashvili}\ \emph {et~al.}(2008)\citenamefont
  {Kahniashvili}, \citenamefont {Kosowsky}, \citenamefont {Gogoberidze},\ and\
  \citenamefont {Maravin}}]{tina2008}%
  \BibitemOpen
  \bibfield  {author} {\bibinfo {author} {\bibfnamefont {T.}~\bibnamefont
  {Kahniashvili}}, \bibinfo {author} {\bibfnamefont {A.}~\bibnamefont
  {Kosowsky}}, \bibinfo {author} {\bibfnamefont {G.}~\bibnamefont
  {Gogoberidze}}, \ and\ \bibinfo {author} {\bibfnamefont {Y.}~\bibnamefont
  {Maravin}},\ }\href {\doibase 10.1103/PhysRevD.78.043003} {\bibfield
  {journal} {\bibinfo  {journal} {PhRvD}\ }\textbf {\bibinfo {volume} {78}},\
  \bibinfo {pages} {043003} (\bibinfo {year} {2008})}\BibitemShut {NoStop}%
\bibitem [{\citenamefont {{Caprini}}\ \emph {et~al.}(2009)\citenamefont
  {{Caprini}}, \citenamefont {{Durrer}},\ and\ \citenamefont
  {{Servant}}}]{Chiara2009}%
  \BibitemOpen
  \bibfield  {author} {\bibinfo {author} {\bibfnamefont {C.}~\bibnamefont
  {{Caprini}}}, \bibinfo {author} {\bibfnamefont {R.}~\bibnamefont {{Durrer}}},
  \ and\ \bibinfo {author} {\bibfnamefont {G.}~\bibnamefont {{Servant}}},\
  }\href {\doibase 10.1088/1475-7516/2009/12/024} {\bibfield  {journal}
  {\bibinfo  {journal} {\jcap}\ }\textbf {\bibinfo {volume} {2009}},\ \bibinfo
  {eid} {024} (\bibinfo {year} {2009})},\ \Eprint
  {http://arxiv.org/abs/0909.0622} {arXiv:0909.0622 [astro-ph.CO]} \BibitemShut
  {NoStop}%
\bibitem [{\citenamefont {Niksa}\ \emph {et~al.}(2018)\citenamefont {Niksa},
  \citenamefont {Schlederer},\ and\ \citenamefont {Sigl}}]{sigl2018}%
  \BibitemOpen
  \bibfield  {author} {\bibinfo {author} {\bibfnamefont {P.}~\bibnamefont
  {Niksa}}, \bibinfo {author} {\bibfnamefont {M.}~\bibnamefont {Schlederer}}, \
  and\ \bibinfo {author} {\bibfnamefont {G.}~\bibnamefont {Sigl}},\ }\href
  {\doibase 10.1088/1361-6382/aac89c} {\bibfield  {journal} {\bibinfo
  {journal} {Class. Quant. Grav.}\ }\textbf {\bibinfo {volume} {35}},\ \bibinfo
  {pages} {144001} (\bibinfo {year} {2018})},\ \Eprint
  {http://arxiv.org/abs/1803.02271} {arXiv:1803.02271 [astro-ph.CO]}
  \BibitemShut {NoStop}%
\bibitem [{\citenamefont {{Sharma}}\ \emph {et~al.}(2020)\citenamefont
  {{Sharma}}, \citenamefont {{Subramanian}},\ and\ \citenamefont
  {{Seshadri}}}]{Sharma+20}%
  \BibitemOpen
  \bibfield  {author} {\bibinfo {author} {\bibfnamefont {R.}~\bibnamefont
  {{Sharma}}}, \bibinfo {author} {\bibfnamefont {K.}~\bibnamefont
  {{Subramanian}}}, \ and\ \bibinfo {author} {\bibfnamefont {T.~R.}\
  \bibnamefont {{Seshadri}}},\ }\href {\doibase 10.1103/PhysRevD.101.103526}
  {\bibfield  {journal} {\bibinfo  {journal} {\prd}\ }\textbf {\bibinfo
  {volume} {101}},\ \bibinfo {eid} {103526} (\bibinfo {year} {2020})},\ \Eprint
  {http://arxiv.org/abs/1912.12089} {arXiv:1912.12089 [astro-ph.CO]}
  \BibitemShut {NoStop}%
\bibitem [{\citenamefont {{Roper Pol}}\ \emph
  {et~al.}(2022{\natexlab{b}})\citenamefont {{Roper Pol}}, \citenamefont
  {{Mandal}}, \citenamefont {{Brandenburg}},\ and\ \citenamefont
  {{Kahniashvili}}}]{RoperPol+21}%
  \BibitemOpen
  \bibfield  {author} {\bibinfo {author} {\bibfnamefont {A.}~\bibnamefont
  {{Roper Pol}}}, \bibinfo {author} {\bibfnamefont {S.}~\bibnamefont
  {{Mandal}}}, \bibinfo {author} {\bibfnamefont {A.}~\bibnamefont
  {{Brandenburg}}}, \ and\ \bibinfo {author} {\bibfnamefont {T.}~\bibnamefont
  {{Kahniashvili}}},\ }\href {\doibase 10.1088/1475-7516/2022/04/019}
  {\bibfield  {journal} {\bibinfo  {journal} {\jcap}\ }\textbf {\bibinfo
  {volume} {2022}},\ \bibinfo {eid} {019} (\bibinfo {year}
  {2022}{\natexlab{b}})},\ \Eprint {http://arxiv.org/abs/2107.05356}
  {arXiv:2107.05356 [gr-qc]} \BibitemShut {NoStop}%
\bibitem [{\citenamefont {{Kahniashvili}}\ \emph {et~al.}(2021)\citenamefont
  {{Kahniashvili}}, \citenamefont {{Brandenburg}}, \citenamefont
  {{Gogoberidze}}, \citenamefont {{Mandal}},\ and\ \citenamefont
  {{Pol}}}]{Kahniashvili+21}%
  \BibitemOpen
  \bibfield  {author} {\bibinfo {author} {\bibfnamefont {T.}~\bibnamefont
  {{Kahniashvili}}}, \bibinfo {author} {\bibfnamefont {A.}~\bibnamefont
  {{Brandenburg}}}, \bibinfo {author} {\bibfnamefont {G.}~\bibnamefont
  {{Gogoberidze}}}, \bibinfo {author} {\bibfnamefont {S.}~\bibnamefont
  {{Mandal}}}, \ and\ \bibinfo {author} {\bibfnamefont {A.~R.}\ \bibnamefont
  {{Pol}}},\ }\href {\doibase 10.1103/PhysRevResearch.3.013193} {\bibfield
  {journal} {\bibinfo  {journal} {PhRvR}\ }\textbf {\bibinfo {volume} {3}},\
  \bibinfo {eid} {013193} (\bibinfo {year} {2021})},\ \Eprint
  {http://arxiv.org/abs/2011.05556} {arXiv:2011.05556 [astro-ph.CO]}
  \BibitemShut {NoStop}%
\bibitem [{\citenamefont {{Dahl}}\ \emph {et~al.}(2021)\citenamefont {{Dahl}},
  \citenamefont {{Hindmarsh}}, \citenamefont {{Rummukainen}},\ and\
  \citenamefont {{Weir}}}]{Jani2021}%
  \BibitemOpen
  \bibfield  {author} {\bibinfo {author} {\bibfnamefont {J.}~\bibnamefont
  {{Dahl}}}, \bibinfo {author} {\bibfnamefont {M.}~\bibnamefont {{Hindmarsh}}},
  \bibinfo {author} {\bibfnamefont {K.}~\bibnamefont {{Rummukainen}}}, \ and\
  \bibinfo {author} {\bibfnamefont {D.}~\bibnamefont {{Weir}}},\ }\href@noop {}
  {\bibfield  {journal} {\bibinfo  {journal} {arXiv e-prints}\ ,\ \bibinfo
  {eid} {arXiv:2112.12013}} (\bibinfo {year} {2021})},\ \Eprint
  {http://arxiv.org/abs/2112.12013} {arXiv:2112.12013 [gr-qc]} \BibitemShut
  {NoStop}%
\bibitem [{\citenamefont {{Roper Pol}}\ \emph
  {et~al.}(2020{\natexlab{b}})\citenamefont {{Roper Pol}}, \citenamefont
  {{Brandenburg}}, \citenamefont {{Kahniashvili}}, \citenamefont {{Kosowsky}},\
  and\ \citenamefont {{Mandal}}}]{RoperPol+20b}%
  \BibitemOpen
  \bibfield  {author} {\bibinfo {author} {\bibfnamefont {A.}~\bibnamefont
  {{Roper Pol}}}, \bibinfo {author} {\bibfnamefont {A.}~\bibnamefont
  {{Brandenburg}}}, \bibinfo {author} {\bibfnamefont {T.}~\bibnamefont
  {{Kahniashvili}}}, \bibinfo {author} {\bibfnamefont {A.}~\bibnamefont
  {{Kosowsky}}}, \ and\ \bibinfo {author} {\bibfnamefont {S.}~\bibnamefont
  {{Mandal}}},\ }\href {\doibase 10.1080/03091929.2019.1653460} {\bibfield
  {journal} {\bibinfo  {journal} {GApFD}\ }\textbf {\bibinfo {volume} {114}},\
  \bibinfo {pages} {130} (\bibinfo {year} {2020}{\natexlab{b}})}\BibitemShut
  {NoStop}%
\bibitem [{\citenamefont {{Brandenburg}}\ \emph {et~al.}(1996)\citenamefont
  {{Brandenburg}}, \citenamefont {{Enqvist}},\ and\ \citenamefont
  {{Olesen}}}]{BEO96}%
  \BibitemOpen
  \bibfield  {author} {\bibinfo {author} {\bibfnamefont {A.}~\bibnamefont
  {{Brandenburg}}}, \bibinfo {author} {\bibfnamefont {K.}~\bibnamefont
  {{Enqvist}}}, \ and\ \bibinfo {author} {\bibfnamefont {P.}~\bibnamefont
  {{Olesen}}},\ }\href {\doibase 10.1103/PhysRevD.54.1291} {\bibfield
  {journal} {\bibinfo  {journal} {\prd}\ }\textbf {\bibinfo {volume} {54}},\
  \bibinfo {pages} {1291} (\bibinfo {year} {1996})},\ \Eprint
  {http://arxiv.org/abs/astro-ph/9602031} {arXiv:astro-ph/9602031 [astro-ph]}
  \BibitemShut {NoStop}%
\bibitem [{\citenamefont {Brandenburg}\ \emph {et~al.}(2017)\citenamefont
  {Brandenburg}, \citenamefont {Kahniashvili}, \citenamefont {Mandal},
  \citenamefont {Roper~Pol}, \citenamefont {Tevzadze},\ and\ \citenamefont
  {Vachaspati}}]{Bran+17}%
  \BibitemOpen
  \bibfield  {author} {\bibinfo {author} {\bibfnamefont {A.}~\bibnamefont
  {Brandenburg}}, \bibinfo {author} {\bibfnamefont {T.}~\bibnamefont
  {Kahniashvili}}, \bibinfo {author} {\bibfnamefont {S.}~\bibnamefont
  {Mandal}}, \bibinfo {author} {\bibfnamefont {A.}~\bibnamefont {Roper~Pol}},
  \bibinfo {author} {\bibfnamefont {A.~G.}\ \bibnamefont {Tevzadze}}, \ and\
  \bibinfo {author} {\bibfnamefont {T.}~\bibnamefont {Vachaspati}},\ }\href
  {\doibase 10.1103/PhysRevD.96.123528} {\bibfield  {journal} {\bibinfo
  {journal} {Phys. Rev. D}\ }\textbf {\bibinfo {volume} {96}},\ \bibinfo
  {pages} {123528} (\bibinfo {year} {2017})}\BibitemShut {NoStop}%
\bibitem [{\citenamefont {{Sharma}}\ and\ \citenamefont
  {{Brandenburg}}()}]{DATA}%
  \BibitemOpen
  \bibfield  {author} {\bibinfo {author} {\bibfnamefont {R.}~\bibnamefont
  {{Sharma}}}\ and\ \bibinfo {author} {\bibfnamefont {A.}~\bibnamefont
  {{Brandenburg}}},\ }\href {\doibase 10.5281/zenodo.7014823} {\bibfield
  {journal} {\bibinfo  {journal} {{Supplemental Material and Datasets for Low
  frequency tail of gravitational wave spectra from hydromagnetic turbulence,
  doi:10.5281/zenodo.7014823 (v2022.08.22)}; see also
  \url{http://norlx65.nordita.org/~brandenb/projects/LowFreqTail/} for easier
  access}\ }10.5281/zenodo.7014823}\BibitemShut {NoStop}%
\bibitem [{\citenamefont {{Caprini}}\ \emph {et~al.}(2004)\citenamefont
  {{Caprini}}, \citenamefont {{Durrer}},\ and\ \citenamefont
  {{Kahniashvili}}}]{CDK04}%
  \BibitemOpen
  \bibfield  {author} {\bibinfo {author} {\bibfnamefont {C.}~\bibnamefont
  {{Caprini}}}, \bibinfo {author} {\bibfnamefont {R.}~\bibnamefont {{Durrer}}},
  \ and\ \bibinfo {author} {\bibfnamefont {T.}~\bibnamefont {{Kahniashvili}}},\
  }\href {\doibase 10.1103/PhysRevD.69.063006} {\bibfield  {journal} {\bibinfo
  {journal} {\prd}\ }\textbf {\bibinfo {volume} {69}},\ \bibinfo {eid} {063006}
  (\bibinfo {year} {2004})},\ \Eprint {http://arxiv.org/abs/astro-ph/0304556}
  {arXiv:astro-ph/0304556 [astro-ph]} \BibitemShut {NoStop}%
\bibitem [{\citenamefont {{Hatori}}(1984)}]{Hat84}%
  \BibitemOpen
  \bibfield  {author} {\bibinfo {author} {\bibfnamefont {T.}~\bibnamefont
  {{Hatori}}},\ }\href {\doibase 10.1143/JPSJ.53.2539} {\bibfield  {journal}
  {\bibinfo  {journal} {JPSJ}\ }\textbf {\bibinfo {volume} {53}},\ \bibinfo
  {pages} {2539} (\bibinfo {year} {1984})}\BibitemShut {NoStop}%
\bibitem [{\citenamefont {{Biskamp}}\ and\ \citenamefont
  {{M{\"u}ller}}(1999)}]{BM99}%
  \BibitemOpen
  \bibfield  {author} {\bibinfo {author} {\bibfnamefont {D.}~\bibnamefont
  {{Biskamp}}}\ and\ \bibinfo {author} {\bibfnamefont {W.-C.}\ \bibnamefont
  {{M{\"u}ller}}},\ }\href {\doibase 10.1103/PhysRevLett.83.2195} {\bibfield
  {journal} {\bibinfo  {journal} {\prl}\ }\textbf {\bibinfo {volume} {83}},\
  \bibinfo {pages} {2195} (\bibinfo {year} {1999})},\ \Eprint
  {http://arxiv.org/abs/physics/9903028} {arXiv:physics/9903028
  [physics.flu-dyn]} \BibitemShut {NoStop}%
\bibitem [{\citenamefont {{Brandenburg}}\ and\ \citenamefont
  {{Kahniashvili}}(2017)}]{BK17}%
  \BibitemOpen
  \bibfield  {author} {\bibinfo {author} {\bibfnamefont {A.}~\bibnamefont
  {{Brandenburg}}}\ and\ \bibinfo {author} {\bibfnamefont {T.}~\bibnamefont
  {{Kahniashvili}}},\ }\href {\doibase 10.1103/PhysRevLett.118.055102}
  {\bibfield  {journal} {\bibinfo  {journal} {PhRvL}\ }\textbf {\bibinfo
  {volume} {118}},\ \bibinfo {eid} {055102} (\bibinfo {year} {2017})},\ \Eprint
  {http://arxiv.org/abs/1607.01360} {arXiv:1607.01360 [physics.flu-dyn]}
  \BibitemShut {NoStop}%
\bibitem [{\citenamefont {{Schekochihin}}(2020)}]{Sch20}%
  \BibitemOpen
  \bibfield  {author} {\bibinfo {author} {\bibfnamefont {A.~A.}\ \bibnamefont
  {{Schekochihin}}},\ }\href@noop {} {\bibfield  {journal} {\bibinfo  {journal}
  {arXiv e-prints}\ ,\ \bibinfo {eid} {arXiv:2010.00699}} (\bibinfo {year}
  {2020})},\ \Eprint {http://arxiv.org/abs/2010.00699} {arXiv:2010.00699
  [physics.plasm-ph]} \BibitemShut {NoStop}%
\bibitem [{\citenamefont {Hosking}\ and\ \citenamefont
  {Schekochihin}(2021)}]{hosking20}%
  \BibitemOpen
  \bibfield  {author} {\bibinfo {author} {\bibfnamefont {D.~N.}\ \bibnamefont
  {Hosking}}\ and\ \bibinfo {author} {\bibfnamefont {A.~A.}\ \bibnamefont
  {Schekochihin}},\ }\href {\doibase 10.1103/PhysRevX.11.041005} {\bibfield
  {journal} {\bibinfo  {journal} {Phys. Rev. X}\ }\textbf {\bibinfo {volume}
  {11}},\ \bibinfo {pages} {041005} (\bibinfo {year} {2021})}\BibitemShut
  {NoStop}%
\bibitem [{\citenamefont {{Zhou}}\ \emph {et~al.}(2022)\citenamefont {{Zhou}},
  \citenamefont {{Sharma}},\ and\ \citenamefont {{Brandenburg}}}]{ZBS22}%
  \BibitemOpen
  \bibfield  {author} {\bibinfo {author} {\bibfnamefont {H.}~\bibnamefont
  {{Zhou}}}, \bibinfo {author} {\bibfnamefont {R.}~\bibnamefont {{Sharma}}}, \
  and\ \bibinfo {author} {\bibfnamefont {A.}~\bibnamefont {{Brandenburg}}},\
  }\href@noop {} {\bibfield  {journal} {\bibinfo  {journal} {arXiv e-prints}\
  ,\ \bibinfo {eid} {arXiv:2206.07513}} (\bibinfo {year} {2022})},\ \Eprint
  {http://arxiv.org/abs/2206.07513} {arXiv:2206.07513 [physics.plasm-ph]}
  \BibitemShut {NoStop}%
\bibitem [{\citenamefont {{Rogachevskii}}\ \emph {et~al.}(2017)\citenamefont
  {{Rogachevskii}}, \citenamefont {{Ruchayskiy}}, \citenamefont {{Boyarsky}},
  \citenamefont {{Fr{\"o}hlich}}, \citenamefont {{Kleeorin}}, \citenamefont
  {{Brandenburg}},\ and\ \citenamefont {{Schober}}}]{Roga_etal17}%
  \BibitemOpen
  \bibfield  {author} {\bibinfo {author} {\bibfnamefont {I.}~\bibnamefont
  {{Rogachevskii}}}, \bibinfo {author} {\bibfnamefont {O.}~\bibnamefont
  {{Ruchayskiy}}}, \bibinfo {author} {\bibfnamefont {A.}~\bibnamefont
  {{Boyarsky}}}, \bibinfo {author} {\bibfnamefont {J.}~\bibnamefont
  {{Fr{\"o}hlich}}}, \bibinfo {author} {\bibfnamefont {N.}~\bibnamefont
  {{Kleeorin}}}, \bibinfo {author} {\bibfnamefont {A.}~\bibnamefont
  {{Brandenburg}}}, \ and\ \bibinfo {author} {\bibfnamefont {J.}~\bibnamefont
  {{Schober}}},\ }\href {\doibase 10.3847/1538-4357/aa886b} {\bibfield
  {journal} {\bibinfo  {journal} {\apj}\ }\textbf {\bibinfo {volume} {846}},\
  \bibinfo {eid} {153} (\bibinfo {year} {2017})},\ \Eprint
  {http://arxiv.org/abs/1705.00378} {arXiv:1705.00378 [physics.plasm-ph]}
  \BibitemShut {NoStop}%
\bibitem [{\citenamefont {{Schober}}\ \emph {et~al.}(2018)\citenamefont
  {{Schober}}, \citenamefont {{Rogachevskii}}, \citenamefont {{Brandenburg}},
  \citenamefont {{Boyarsky}}, \citenamefont {{Fr{\"o}hlich}}, \citenamefont
  {{Ruchayskiy}},\ and\ \citenamefont {{Kleeorin}}}]{Schober+18}%
  \BibitemOpen
  \bibfield  {author} {\bibinfo {author} {\bibfnamefont {J.}~\bibnamefont
  {{Schober}}}, \bibinfo {author} {\bibfnamefont {I.}~\bibnamefont
  {{Rogachevskii}}}, \bibinfo {author} {\bibfnamefont {A.}~\bibnamefont
  {{Brandenburg}}}, \bibinfo {author} {\bibfnamefont {A.}~\bibnamefont
  {{Boyarsky}}}, \bibinfo {author} {\bibfnamefont {J.}~\bibnamefont
  {{Fr{\"o}hlich}}}, \bibinfo {author} {\bibfnamefont {O.}~\bibnamefont
  {{Ruchayskiy}}}, \ and\ \bibinfo {author} {\bibfnamefont {N.}~\bibnamefont
  {{Kleeorin}}},\ }\href {\doibase 10.3847/1538-4357/aaba75} {\bibfield
  {journal} {\bibinfo  {journal} {\apj}\ }\textbf {\bibinfo {volume} {858}},\
  \bibinfo {eid} {124} (\bibinfo {year} {2018})}\BibitemShut {NoStop}%
\bibitem [{\citenamefont {{Pencil Code Collaboration}}\ \emph
  {et~al.}(2021)\citenamefont {{Pencil Code Collaboration}}, \citenamefont
  {{Brandenburg}}, \citenamefont {{Johansen}}, \citenamefont {{Bourdin}},
  \citenamefont {{Dobler}}, \citenamefont {{Lyra}}, \citenamefont
  {{Rheinhardt}}, \citenamefont {{Bingert}}, \citenamefont {{Haugen}},
  \citenamefont {{Mee}}, \citenamefont {{Gent}}, \citenamefont {{Babkovskaia}},
  \citenamefont {{Yang}}, \citenamefont {{Heinemann}}, \citenamefont
  {{Dintrans}}, \citenamefont {{Mitra}}, \citenamefont {{Candelaresi}},
  \citenamefont {{Warnecke}}, \citenamefont {{K{\"a}pyl{\"a}}}, \citenamefont
  {{Schreiber}}, \citenamefont {{Chatterjee}}, \citenamefont
  {{K{\"a}pyl{\"a}}}, \citenamefont {{Li}}, \citenamefont {{Kr{\"u}ger}},
  \citenamefont {{Aarnes}}, \citenamefont {{Sarson}}, \citenamefont {{Oishi}},
  \citenamefont {{Schober}}, \citenamefont {{Plasson}}, \citenamefont
  {{Sandin}}, \citenamefont {{Karchniwy}}, \citenamefont {{Rodrigues}},
  \citenamefont {{Hubbard}}, \citenamefont {{Guerrero}}, \citenamefont
  {{Snodin}}, \citenamefont {{Losada}}, \citenamefont {{Pekkil{\"a}}},\ and\
  \citenamefont {{Qian}}}]{JOSS}%
  \BibitemOpen
  \bibfield  {author} {\bibinfo {author} {\bibnamefont {{Pencil Code
  Collaboration}}}, \bibinfo {author} {\bibfnamefont {A.}~\bibnamefont
  {{Brandenburg}}}, \bibinfo {author} {\bibfnamefont {A.}~\bibnamefont
  {{Johansen}}}, \bibinfo {author} {\bibfnamefont {P.}~\bibnamefont
  {{Bourdin}}}, \bibinfo {author} {\bibfnamefont {W.}~\bibnamefont {{Dobler}}},
  \bibinfo {author} {\bibfnamefont {W.}~\bibnamefont {{Lyra}}}, \bibinfo
  {author} {\bibfnamefont {M.}~\bibnamefont {{Rheinhardt}}}, \bibinfo {author}
  {\bibfnamefont {S.}~\bibnamefont {{Bingert}}}, \bibinfo {author}
  {\bibfnamefont {N.}~\bibnamefont {{Haugen}}}, \bibinfo {author}
  {\bibfnamefont {A.}~\bibnamefont {{Mee}}}, \bibinfo {author} {\bibfnamefont
  {F.}~\bibnamefont {{Gent}}}, \bibinfo {author} {\bibfnamefont
  {N.}~\bibnamefont {{Babkovskaia}}}, \bibinfo {author} {\bibfnamefont {C.-C.}\
  \bibnamefont {{Yang}}}, \bibinfo {author} {\bibfnamefont {T.}~\bibnamefont
  {{Heinemann}}}, \bibinfo {author} {\bibfnamefont {B.}~\bibnamefont
  {{Dintrans}}}, \bibinfo {author} {\bibfnamefont {D.}~\bibnamefont {{Mitra}}},
  \bibinfo {author} {\bibfnamefont {S.}~\bibnamefont {{Candelaresi}}}, \bibinfo
  {author} {\bibfnamefont {J.}~\bibnamefont {{Warnecke}}}, \bibinfo {author}
  {\bibfnamefont {P.}~\bibnamefont {{K{\"a}pyl{\"a}}}}, \bibinfo {author}
  {\bibfnamefont {A.}~\bibnamefont {{Schreiber}}}, \bibinfo {author}
  {\bibfnamefont {P.}~\bibnamefont {{Chatterjee}}}, \bibinfo {author}
  {\bibfnamefont {M.}~\bibnamefont {{K{\"a}pyl{\"a}}}}, \bibinfo {author}
  {\bibfnamefont {X.-Y.}\ \bibnamefont {{Li}}}, \bibinfo {author}
  {\bibfnamefont {J.}~\bibnamefont {{Kr{\"u}ger}}}, \bibinfo {author}
  {\bibfnamefont {J.}~\bibnamefont {{Aarnes}}}, \bibinfo {author}
  {\bibfnamefont {G.}~\bibnamefont {{Sarson}}}, \bibinfo {author}
  {\bibfnamefont {J.}~\bibnamefont {{Oishi}}}, \bibinfo {author} {\bibfnamefont
  {J.}~\bibnamefont {{Schober}}}, \bibinfo {author} {\bibfnamefont
  {R.}~\bibnamefont {{Plasson}}}, \bibinfo {author} {\bibfnamefont
  {C.}~\bibnamefont {{Sandin}}}, \bibinfo {author} {\bibfnamefont
  {E.}~\bibnamefont {{Karchniwy}}}, \bibinfo {author} {\bibfnamefont
  {L.}~\bibnamefont {{Rodrigues}}}, \bibinfo {author} {\bibfnamefont
  {A.}~\bibnamefont {{Hubbard}}}, \bibinfo {author} {\bibfnamefont
  {G.}~\bibnamefont {{Guerrero}}}, \bibinfo {author} {\bibfnamefont
  {A.}~\bibnamefont {{Snodin}}}, \bibinfo {author} {\bibfnamefont
  {I.}~\bibnamefont {{Losada}}}, \bibinfo {author} {\bibfnamefont
  {J.}~\bibnamefont {{Pekkil{\"a}}}}, \ and\ \bibinfo {author} {\bibfnamefont
  {C.}~\bibnamefont {{Qian}}},\ }\href {\doibase 10.21105/joss.02807}
  {\bibfield  {journal} {\bibinfo  {journal} {JOSS}\ }\textbf {\bibinfo
  {volume} {6}},\ \bibinfo {eid} {2807} (\bibinfo {year} {2021})}\BibitemShut
  {NoStop}%
\end{thebibliography}%
\end{document}